# $^{57}$Fe Mössbauer study of unusual magnetic structure of multiferroic 3$R$-AgFeO$_2$

Alexey V. Sobolev[1], Vyacheslav S. Rusakov[2], Alexander S. Moskvin[3], Alexei M. Gapochka[2], Alexei A. Belik[4], Iana S. Glazkova[1], Gerard Demazeau[5], Igor A. Presniakov[1]

[1] *Department of Chemistry, Lomonosov Moscow State University, 119991 Moscow, Russia*

[2] *Department of Physics, Lomonosov Moscow State University, 119991 Moscow, Russia*

[3] *Department of Theoretical Physics, Institute of Natural Sciences, Ural Federal University, 620083 Ekaterinburg, Russia*

[4] *Research Center for Functional Materials, National Institute for Materials Science (NIMS), 1-1 Namiki, Tsukuba, Ibaraki 305-0044, Japan*

[5] *HPBioTECH, 22 rue Denis Papin, ZI de La Rivière 33850 LEOGNAN, France*

We report new results of a $^{57}$Fe Mössbauer study of multiferroic 3$R$-AgFeO$_2$ powder samples performed in a wide temperature range, including two points, $T_{N1} \approx 14$ K and $T_{N2} \approx 9$ K, of magnetic phase transitions. At the intermediate temperature range, $T_{N2} < T < T_{N1}$, the $^{57}$Fe Mössbauer spectra can be described in terms of collinear spin-density-waves (SDW) with the inclusion of many high-order harmonics, indicating that the real magnetic structure of this ferrite appears to be more complicated than a pure sinusoidally modulated SDW. The spectra at low temperatures, $T < T_{N2}$, consist of a Zeeman pattern with line broadenings and sizeable spectral asymmetry. It has been shown that the observed spectral shape is consistent with a transition to the elliptical cycloidal magnetic structure. An analysis of the experimental spectra was carried out under the assumption that the electric hyperfine interactions are modulated when the Fe$^{3+}$ magnetic moment rotates with respect to the principal axis of the EFG tensor and emergence of the strong anisotropy of the magnetic hyperfine field $H_{hf}$ at the $^{57}$Fe nuclei. The large and temperature-independent anharmonicity parameter, $m \approx 0.78$, of the cycloidal spin structure obtained from the experimental spectra results from easy-axis anisotropy in the plane of rotation of the iron spin. Analysis of different mechanisms of spin and hyperfine interactions in 3$R$-AgFeO$_2$ and its structural analogue CuFeO$_2$ points to a specific role played by the topology of the exchange coupling and the oxygen polarization in the delafossite structures.

# INTRODUCTION

$A M O_2$ delafossite-like oxides ($A$ = Cu$^+$, Ag$^+$ and $M$ = Cr$^{3+}$, Fe$^{3+}$, Co$^{3+}$, Ni$^{3+}$) with triangular, geometrically frustrated spin structures attract much attention as magnetoelectric multiferroic materials [1-7]. The ferroelectricity in this class of multiferroics appears as a result of the phase transition, inducing an unusual magnetic structure that breaks the crystal symmetry. Geometrical spin frustration is one of the main origins of such a magnetic state [4]. The frustrated system, due to the vast degeneracy arising from competing magnetic interactions of transition ions, often displays non-centrosymmetric noncollinear or long-period-modulated collinear orders.

It would be useful to compare the local structural and magnetic properties of CuFeO$_2$ and 3$R$-AgFeO$_2$ oxides containing identical magneto-active Fe$^{3+}$ ions but with different diamagnetic Cu$^+$ and Ag$^+$ cations. Both of the oxides have a delafossite-like structure with the rhombohedral space group $R\bar{3}m$ at room temperature. The structure consists of Fe$^{3+}$ hexagonal layers along the $c$ axis, which are separated by nonmagnetic $A^+$-O$^{2-}$ dumbbells ($A$ = Cu, Ag) (Fig. 1a). A decrease in temperature causes a symmetry lowering from the $R\bar{3}m$ space group to the monoclinic $C2/m$ [9]. According to magnetic data for CuFeO$_2$ [10] and our earlier studies of 3$R$-AgFeO$_2$ [11], these oxides exhibit two successive magnetic transitions at $T_{N1} \approx 14$ K, $T_{N2} \approx 11$ K (Cu) and $T_{N1} \approx 14$ K, $T_{N2} \approx 9$ K (Ag). Below $T_{N1}$, the oxides become magnetically ordered, with a sinusoidally modulated and partially disordered structure indexed by incommensurate propagation vectors $Q = 2\pi(0, q, \frac{1}{2})_m$ (Cu) [10] and $Q = 2\pi(-\frac{1}{2}, q, \frac{1}{2})_m$ [12], respectively, with the wave number $q \sim \frac{2}{5}$ depending on temperature.

Despite the above similarity of the structural and thermodynamic parameters of CuFeO$_2$ and 3$R$-AgFeO$_2$, the magnetic ordering in these systems clearly demonstrates the difference in their magnetic ground state at low temperatures, $T < T_{N2}$. The copper ferrite has a collinear four-sublattice (4SL) ground state ↑↑↓↓ with a commensurate propagation vector $Q = (0, \frac{1}{2}, \frac{1}{2})_m$ [10] in the monoclinic cell. At the same time, according to recent neutron diffraction experiments [9, 12], the magnetic ordering in 3$R$-AgFeO$_2$ at $T \leq T_{N2}$ is in the form of an elliptical cycloid with an incommensurate propagation wave vector $Q = 2\pi(-\frac{1}{2}, q, \frac{1}{2})_m$, with $q \approx 0.2026$.

Such an essential difference in the nonpolar commensurate state of CuFeO$_2$ and the polar magnetic structure of 3$R$-AgFeO$_2$ clearly underlines that the nonmagnetic $A$ (= Cu$^+$, Ag$^+$) ions play a crucial role in the magnetic exchange interactions and multiferroic behavior of the delafossite $A$FeO$_2$ compounds [9, 13]. The ground state spin structure and its temperature variation in the quasi-2$D$ systems $A$FeO$_2$ is far from trivial, due to a competition between several interactions. First of all these are the intra-layer ($J_{ij}^{(\text{intra})}$) and inter-layer ($J_{ij}^{(\text{inter})}$) isotropic exchange coupling described by the Hamiltonian:

$$\hat{H} = \sum_{i,j} J_{ij}^{(\text{inter})} \mathbf{S}_i \cdot \mathbf{S}_j + \sum_{i,j} J_{ij}^{(\text{intra})} \mathbf{S}_i \cdot \mathbf{S}_j, \quad (1)$$

where $J_{ij}^{(\text{inter})} \equiv J_0$ and $J_{ij}^{(\text{intra})} \equiv J_{1,2,3}$ (Fig. 1b). The intra-layer exchange interactions in $3R$-AgFeO$_2$ and CuFeO$_2$ are significantly affected by the substitution of the nonmagnetic $A$-site cations in spite of the common low-temperature monoclinic symmetry in both oxides. We argue that this effect can be explained to be a typical one for the edge-shared exchange-coupled clusters. Indeed, the antiferromagnetic kinetic contribution to the superexchange integral Fe$^{3+}$($d^5$, $^6A_{1g}$)-O$^{2-}$-Fe$^{3+}$($d^5$, $^6A_{1g}$) can be written as follows [14]:

$$J_{\text{FeFe}} = \frac{2}{25U}\left[(t_{ss} + t_{\sigma\sigma}\cos\theta)^2 + t_{\sigma\pi}^2 \sin^2\theta + t_{\pi\pi}^2(2 - \sin^2\theta)\right] > 0, \quad (2)$$

where $\theta$ is the Fe-O-Fe bonding angle, $t_{\sigma\sigma} > t_{\sigma\pi} > t_{\pi\pi} > t_{s\sigma}$ are positive definite $d$-$d$ transfer integrals, $U$ is a mean $d$-$d$ transfer energy (correlation energy). For the edge-shared FeO$_6$ octahedra with the bonding angle close to 90° as in delafossite structure the strongest $\sigma$-$\sigma$ bond is invalidated and the weakened antiferromagnetic contribution starts to compete with a ferromagnetic ($J_{\text{pot}} < 0$) potential (Heisenberg) exchange giving rise to a striking sensitivity of the net exchange integral on rather small changes of structural parameters, such as superexchange bonding angles and cation-anion separations. Furthermore, the compensation effect does promote the relative role of next-nearest-neighbors (*nnn*) Fe-O-Fe and next-next-nearest-neighbors (*nnnn*) Fe-O-O-Fe interactions (Fig. 1b).

Another characteristic feature of the topology of exchange-interactions in delafossites $A$FeO$_2$ is that, different from the most part of ferrites, an O$^{2-}$ ion belongs to three Fe-O-Fe bonds that makes the exchange coupling to be extremely sensitive to oxygen displacements and its electric polarization thus providing paths for understanding the exotic spin-lattice coupling phenomena, specifically spin-driven bond order, in geometrically frustrated magnets [15]. The comprehensive analysis of the isotropic superexchange coupling in delafossites has to take into account a strong electric polarization of the intermediate oxygen ions.

Furthermore, specific spin structures in $3R$-AgFeO$_2$ and CuFeO$_2$ are related with usually more weak anisotropic interactions such as a single-ion anisotropy (SIA) and Dzyaloshinskii-Moriya (DM) coupling:

$$E_{SIA} + E_{DM} = D\sum_{i=1} S_{iz}^2 + \sum_{i>j}(\mathbf{d}_{ij} \cdot [\mathbf{S}_i \times \mathbf{S}_j]), \quad (3)$$

where $D$ is anisotropy parameter, $\mathbf{d}_{ij}$ is the axial Dzyaloshinskii vector. Within a linear approximation on the noncubic distortions the energy of the single-ion anisotropy in FeO$_6$ octahedral clusters can be represented in the main coordinate system (O$_{x,y,z}$ || $C_4$) as follows:

$$E_{SIA} = \sum_i k_i \alpha_i^2 + \sum_{i<j} k_{ij} \alpha_i \alpha_j \quad (4)$$

where $\alpha_i$ is direction cosine of the $S_i$ spin. The anisotropy constants are proportional to components of the octahedron deformation tensor: $k_i = b_E \epsilon_{ii}$, $k_{ij} = b_{T2}\epsilon_{ij}$, where $\epsilon_{ii} = (l_i - l_0)/l_0$, $\epsilon_{ij} = ½(\pi/2 - \theta_{ij})$, $l_i$ is a cation – $i$-th ligand separation, $l_0 = ⅓\sum l_i$ is a mean cation-anion separation, $\theta_{ij}$ is the $i$-th ligand – cation – $j$-th ligand bond angle. Magnetoelastic parameters ($b_E, b_{T2}$) that determine the contribution of the distortions such as elongation-contraction of the cation-anion bonds and deviations of the anion-cation-anion bonding angles from $\pi/2$, respectively, were estimated earlier for FeO$_6$ clusters in $R$FeO$_3$: $b_E \approx 24$ cm$^{-1}$; $b_{T2} \approx 6$ cm$^{-1}$ [16]. The positive sign of the $b_{T2}$ produces an important result that the trigonal distortion along $C_3$ axis with the Fe$^{3+}$O$_6$ octahedron contraction along the axis ($\theta_{ij} > \pi/2$) makes the axis to be the easy one, while the octahedron elongation along the $C_3$ axis makes the respective perpendicular plane (111) to be the easy plane. For 3$R$-AgFeO$_2$ and CuFeO$_2$ with the FeO$_6$ octahedrons contracted along hexagonal $c$-axis we arrive at an easy-axis type of the single-ion anisotropy with a rather large value of the anisotropy parameter $D$:

$$D \approx \frac{4}{25} \cdot \frac{\pi}{360^0}(90^0 - 96.6^0) \cdot 6 (\text{cm}^{-1}) \approx -0.008 (\text{meV}), \qquad (5)$$

where we used experimental data for the trigonal distortion of the Fe$^{3+}$O$_6$ octahedra in 3$R$-AgFeO$_2$ at $T > T_{N1}$ : $\theta_{ij} = 96.6°$ [12]. Interestingly, this estimate is close to experimentally found $D = -0.01$ meV in CuFe$_{1-x}$Ga$_x$O$_2$ (x = 0.035) [17]. However, the Fe$^{3+}$ single-ion anisotropy in CuFeO$_2$ can be as large as $D \approx -0.2$ meV with a puzzling downfall to $D \approx -0.01$ meV under a slight substitution Fe for Ga$^{3+}$ ions [17]. One of the most probable explanations for this unexpected anisotropy can be related with a markedly large and sensitive electric polarization of oxygen ions [18, 19] which seems to be a specific property of delafossite structure as compared with many other oxides. The electric field induced O($sp$)-hybridization accompanied with an effective off-center shift of the valence O2$p$-shell can result in a large orbital anisotropy through the anisotropic Fe 3$d$-O(2$p$) hybridization.

For so-called $S$-type 3$d$-ions (Mn$^{2+/4+}$, Fe$^{3+}$, Cr$^{3+}$, Ni$^{2+}$) with orbitally non-degenerated ground state one might use a simple formula for the Dzyaloshinskii vector for a cation(1)-anion-cation(2) superexchange bond [20-22]:

$$\boldsymbol{d}(\theta) = d_{12}(\cos\theta) \cdot [\boldsymbol{r}_1 \times \boldsymbol{r}_2] = \frac{1}{2l^2}d_{12}(\cos\theta) \cdot [\boldsymbol{R}_{12} \times \boldsymbol{\rho}_{12}], \qquad (6)$$

where $\boldsymbol{R}_{12} = \boldsymbol{R}_1 - \boldsymbol{R}_2$, $\boldsymbol{\rho}_{12} = \boldsymbol{R}_1 + \boldsymbol{R}_2$, where $\boldsymbol{R}_1, \boldsymbol{R}_2$ are anion-cation(1,2) bond vectors, and $\boldsymbol{\rho}_{12} \perp \boldsymbol{R}_{12}$, $\boldsymbol{r}_1 = \boldsymbol{R}_1/l$, $\boldsymbol{r}_2 = \boldsymbol{R}_2/l$ are respective unit vectors. The sign of the scalar parameter $d_{12}(\cos\theta)$ can be addressed to be the sign of the Dzyaloshinskii vector.

Hereafter we do not consider microscopic mechanisms of Dzyaloshinskii-Moriya coupling to be supposedly a main source of the helical ordering and multiferroicity in $A$FeO$_2$ delafossites and shall limit ourselves to a simplified continual Landau-Ginzburg approximation for main magnetic interactions that provides nonetheless a comprehensive description of the helical ordering in delafossites. The Landau-Ginzburg free energy ($f_{LG}$) density of the $A$FeO$_2$ systems can be written as follows [23]:

$$f_{LG} = \sum_i A_i (\nabla S_i)^2 - \alpha \mathbf{P}[\mathbf{S}(\nabla \mathbf{S}) - (\mathbf{S}\nabla)\mathbf{S}] + DS_z^2, \qquad (7)$$

where the first term is an exchange interaction with an exchange stiffness constants $A_i$ ($i = x, y, z$); the second term is the Lifshitz invariant form of the specific Dzyaloshinskii-Moriya coupling, $\mathbf{P}$ is the spontaneous electric polarization vector, $\alpha$ is the inhomogeneous magnetoelectric interaction constant; the third term is a magnetic anisotropy term with $D$ to be the uniaxial magnetic anisotropy constant. For classical spins $\mathbf{S} = S_0(\sin\vartheta\cos\varphi, \sin\vartheta\sin\varphi, \cos\vartheta)$, the minimization of the free-energy functional $F = \int f_{LG} dV$ by the Lagrange-Euler method gives for the functions $\varphi(x,y,z)$ and $\vartheta(x,y,z)$ [23] as follows:

$$\cos\vartheta(x) = \operatorname{sn}[(\pm 4\mathrm{K}(m)/\lambda)x, m], \qquad (8)$$

with $\varphi = const$ (ignoring the small canting out of the cycloid plane), that is, an anharmonic cycloid in which the spin component $S_z$ along $z\|V_{zz}$ direction is given by the elliptic Jacobi function sn(...), where $\lambda$ is the period of the cycloid, $\mathrm{K}(m) = \int_0^{\pi/2} d\vartheta/(1 - m\cos^2\vartheta)^{1/2}$ is the complete elliptic integral of the first kind, and $m$ is the anharmonicity parameter related to the distortion (anharmonicity) of the spiral structure. The classical distortion of the spiral corresponds to a redistribution of the spin vectors around a circle. The value of $m$ is related to the uniaxial anisotropy $D$ as follows: when $m \to 0$ and $D = 0$, all spin directions are equivalent, giving a simple spiral that contains only one turn angle, $\Delta\vartheta_0$. When $m \to 1$ and $D < 0$, the spin rotation is no longer isotropic, even in the classical limit, and the spins favor the easy axis direction [23].

The *Lifshitz* term in (8) can be easily reproduced within so-called "spin-current" mechanism of the spin-dependent electric polarization [KNB] when:

$$\mathbf{P} \propto \sum_{ij} \left[ \mathbf{R}_{ij} \times [\mathbf{S}_i \times \mathbf{S}_j] \right]. \qquad (9)$$

"Ferroelectricity caused by spin-currents" has established itself as one of the leading paradigms for both theoretical and experimental investigations in the field of a strong multiferroic coupling. However, the spin-current model cannot explain an emergence of ferroelectricity associated with proper crew magnetic ordering in several multiferroics, including CuFeO$_2$, and specific features of the electric polarization induced by the cycloid structure in 3$R$-AgFeO$_2$ [9]. Indeed, the model stems somehow or other from exchange-relativistic effect, or Dzyaloshinskii-Moriya coupling, however, it does not take into account specific effects of the superexchange geometry. According to microscopic theory by *Moskvin et al.* [24, 25] the relativistic spin-dependent electric polarization for cation(1)-anion-cation(2) system can be written as follows:

$$\boldsymbol{P}_{12}^{\text{rel}} = -\frac{1}{J_{12}} \Pi_{12} \left( d_{12} \cdot [\boldsymbol{s}_1 \times \boldsymbol{s}_2] \right), \qquad (10)$$

where $J_{12}$ is a superexchange integral, $\Pi_{12}$ is a so-called exchange-dipole moment which in general can be written as a superposition of the "longitudinal" and "transversal" contributions:

$$\boldsymbol{\Pi}_{12} = p_\parallel \boldsymbol{R}_{12} + p_\perp \boldsymbol{\rho}_{12}, \qquad (11)$$

where $p_\parallel$ does not vanish only for crystallographically nonequivalent cations. In other words, the exchange-relativistic contribution to the dipole moment $\boldsymbol{P}_{12}^{\text{rel}}$ is a superposition of the "longitudinal" and "transversal" mutually orthogonal contributions determined by the superexchange geometry, while the "spin-current" factor $[\boldsymbol{s}_1 \times \boldsymbol{s}_2]$ does only modulate its value.

Numerous experimental findings on the electric field effect in ESR and their theoretical studies [26] can be used for a direct estimation of the single-ion contribution to the ME coupling in different multiferroics [25]. Indeed, the single-ion spin-dependent electric dipole moment widely used for many years in this field is

$$d_i = -\tfrac{1}{2} R_{ijk} \{S_j, S_k\} \qquad (12)$$

and for $R_{ijk}$ parameters one finds numerous experimental data [26]. It should be noted that if $R_{ijk} = \delta_{ij} R_k$, where vector $\boldsymbol{R}$ is the spin-spin bonding vector, we arrive at the expression for the spin-dependent dipole moment introduced by *Arima* [8] who made use of the term for explanation of the magnetoelectric polarization in $CuFeO_2$ with its proper-screw spin order.

The above physical properties, including unusual magnetic structure and magnetism, primarily depend on peculiarities of the electronic structure and crystal local surrounding of iron ions. Thus, $^{57}$Fe Mössbauer spectroscopy is one of the most powerful local methods for studying the $A$FeO$_2$ compounds. Since the local magnetization of the iron ions induces a hyperfine field ($H_{\text{hf}}$) at $^{57}$Fe nuclei proportional to the local amplitude of the SDW via the core spin polarization mechanism, Mössbauer spectroscopy could be very useful to study the static magnetic order and low-energy spin fluctuations. Furthermore, since the basic mechanisms of the magnetic hyperfine interactions are in many respect analogous to that accepted in the theory of magnetic exchange interactions, studies of various contributions to the experimental $H_{\text{hf}}$ value can be quite useful in determining the relative importance of the various mechanisms of spin transfer within the Fe-O-Fe and Fe-Fe bonds in the $A$FeO$_2$ structures.

While $CuFeO_2$ has been extensively investigated by Mössbauer spectroscopy [27-32], $AgFeO_2$ has been preliminarily studied only in our works [33, 34]. $^{57}$Fe Mössbauer measurements performed for 3$R$-AgFeO$_2$ in the paramagnetic temperature region $T > T_{\text{N1}}$ showed that all iron cations occupy unique crystallographic positions, in accordance with the crystal data [12]. Moreover, a self-consistent analysis of the complex magnetic hyperfine spectra at $T < T_{\text{N2}}$ [33] has been proposed. A reasonable fit was obtained by using the quasi-continuous variation of the hyperfine magnetic field ($H_{hf}$) amplitude, with the iron spin orientation varying in the (*bc*) plane. It was shown [33, 34] that to obtain a good fit, an anharmonicity (bunching) of the iron spins along certain directions in the (*bc*) plane is necessary. However, we did not discuss the origin of such anharmonicity in the context of the electronic structure of iron ions in the ferrite. Moreover, there was no information on the temperature evolution of the hyperfine Zeeman structure of Mössbauer spectra, in particular, at $T_{N2} < T < T_{N1}$ and near the critical point ($T \approx T_{N1}$), which would be very useful to clarify the nature of the magnetic phase transition.

In this work, we present the results of a detailed Mössbauer study of the ferrite 3*R*-AgFeO$_2$ in a wide range of temperatures, including both magnetic phase transitions ($T_{N1}$ and $T_{N2}$). The shape of the $^{57}$Fe Mössbauer spectra is shown to undergo significant changes when the type and character of spin ordering of the iron sublattice are changed. At $T_{N2} < T < T_{N1}$, the magnetically split spectra are fitted in terms of incommensurate spin density wave (IC-SDW) modulation [12]. In the range $T < T_{N2}$, the $^{57}$Fe Mössbauer spectra are analyzed assuming a space-modulated cycloidal magnetic structure. Such an approach allows us to reproduce, from experimental spectra, the profile of the spatial anisotropy of the hyperfine field, $H_{hf}$. In addition, we carried out a detailed analysis of the temperature dependences of hyperfine parameters, and a discussion is provided in light of the peculiarities of the electronic and magnetic states of the iron ions in 3*R*-AgFeO$_2$. The obtained data are compared with earlier published Mössbauer data for the CuFeO$_2$ [27-32] and BiFeO$_3$ [35, 36] also revealing multiferroic properties.

## EXPERIMENT

3*R*-AgFeO$_2$ was synthesized from a stoichiometric mixture of Ag$_2$O (99.99 %) and Fe$_2$O$_3$ (99.999%). The mixture was placed in Au capsules and treated at 3 GPa and 1073-1173 K for 2 h (heating time to the desired temperatures was 10 min) in a belt-type high-pressure apparatus. After the heat treatments, the samples were quenched to room temperature (RT), and the pressure was slowly released. The samples were black dense pellets.

X-ray powder diffraction (XRPD) data were collected at RT on a RIGAKU MiniFlex600 diffractometer using CuK$\alpha$ radiation (2$\theta$ range of 10–80°, a step width of 0.02°, and scan speed of 1 deg/min). 3*R*-AgFeO$_2$ samples contained small amounts of Fe$_2$O$_3$ impurities.

Magnetic susceptibilities ($\chi = \mathbf{M}/\mathbf{H}$) were measured using a SQUID magnetometer (Quantum Design, MPMS-1T) between 2 and 300 K in different applied magnetic fields under both zero-field-cooled (ZFC) and field-cooled on cooling (FCC) conditions. No dependence on the applied fields was observed, and no difference between ZFC and FCC curves was detected. Isothermal magnetization measurements were performed at 5 K between -1 T and 1 T; the M-H curves were linear with the magnetization of $0.0455\mu_B$ at 5 K and 1 T. Specific heat, $C_p$, was recorded from 60 K to 2 K at 0 and 90 kOe by a pulse relaxation method using a commercial calorimeter (Quantum Design PPMS).

The $^{57}$Fe Mössbauer spectra were recorded at 4.6 – 300 K using a conventional constant-acceleration spectrometer. The radiation source $^{57}$Co(Rh) was kept at room temperature (RT). All isomer shifts refer to the α-Fe at RT. The experimental spectra were processed and analyzed using methods of the model fitting and reconstruction of the distribution of the hyperfine parameters corresponding to partial spectra implemented in the *SpectrRelax* program [37].

## RESULTS AND DISCUSSION

### 1. X-ray Diffraction data

The XRPD patterns of the synthesized samples showed the formation of the rhombohedral 3*R*-AgFeO$_2$ phase (space group *R*-3*m*) without traces of the hexagonal 2*H*-AgFeO$_2$ phase (space group *P*6$_3$/*mmc*) that differs from 3*R*-AgFeO$_2$ by the stacking of the (FeO$_2$) planes [2, 11]. The refined lattice parameters of 3*R*-AgFeO$_2$ ($a = 3.0386(1)$ Å and $c = 18.5844(4)$ Å) are in good agreement with literature data [11, 12]. In what follows, the rhombohedral 3*R*-AgFeO$_2$ phase will be referred without "3*R*".

### 2. Magnetic and thermodynamic data

A discussion of the magnetic and thermodynamic properties of the AgFeO$_2$ ferrite was given in our previous paper [11] and also in a recent, detailed work [9, 12]. Here, we only present the basic magnetic parameters that characterize the quality of the samples under investigation and the data necessary for further discussion of the $^{57}$Fe Mössbauer spectra of AgFeO$_2$.

The temperature-dependence of the static magnetic susceptibility ($\chi$), measured in a field of $H_{ext} = 1$ kOe, is plotted in Fig. 2a. At high temperatures, 70 K < *T* < 345 K, the inverse susceptibility $\chi^{-1}(T)$ follows the Curie-Weiss law $(T - \Theta_{CW})/C_m$ (insert for Fig. 2a). The obtained Curie constant, $C_m = 2.73(3)$ cm$^3$·K/mol, yields an effective magnetic moment $\mu_{eff} = 2.67 \cdot (C_m)^{1/2} \approx 6.06(3)$ $\mu_B$, which is slightly higher than the spin-only value $2\sqrt{S(S+1)}\mu_B$ expected for the high-spin Fe$^{3+}$ ions with $S = 5/2$ and quenched orbital moment ($<L> \approx 0$). The small discrepancy may be related to the partially unquenched orbital moment for the half-filled $d^5$ system, making the effective $g_{eff}$-factor slightly higher than 2 [38].

The experimental Curie-Weiss temperature $\Theta_{CW} = -139(2)$ K appears to be significantly higher than the critical point $T_{N1} \approx 14(1)$ K of the long-range magnetic transition, deduced from specific-heat measurements $C_m(T)$ (Fig. 2b). This finding reveals the presence of strong frustration, indicated by the high value of the ratio $|\Theta_{CW}/T_{N1}| \approx 10$, which is in complete agreement with the triangular cationic topology of the delafossite-like lattice. It should be noted the obtained ratio is higher than the $|\Theta_{CW}/T_{N1}| \approx 5$ value for $CuFeO_2$ [10], indicating that the $AgFeO_2$ appears to be more frustrated. In the mean-field approximation, the Curie-Weiss temperature is related to the exchange parameters by the relation $|\Theta_{CW}| = \alpha \cdot \sum_i z_i J_i$, where $\alpha \equiv 2S(S+1)/3k_B$, and $z_i$ is the number of nearest neighbors of iron connected by exchange intralayer ($J_i$) and interlayer ($J_0$) coupling [39]. The Neel temperature is given in the form $T_N = \alpha J(q)$, where $J(q)$ is the Fourier transform of the exchange integral $J(r_i)$, and $q$ is the propagation vector representing the magnetic structure. Taking into account the almost complete coincidence of the $T_{N1}$ values of $CuFeO_2$ ($T_{N1} = 16$ K, [9]) and $AgFeO_2$ ($T_{N1} = 14(1)$ K, Fig. 2), the observed increase of frustration ($\Theta_{CW}/T_N$) in the case of $AgFeO_2$ may be ascribed to the strengthening of its own magnetic interactions.

The above results clearly demonstrate that the microscopic mechanisms for lifting the magnetic frustration within the $AgFeO_2$ and $CuFeO_2$ lattices, both of which have a common low-temperature monoclinic structure ($T < T_{N1}$), could be very different. A similar conclusion was recently drawn in the comparative analysis of the lattice distortion in these ferrites [13]. In particular, it has been shown that the $b_m$ axis in the monoclinic basis contracts at $T < T_{N1}$ in $AgFeO_2$ and elongates in $CuFeO_2$. It is important to note that all the known changes in the spin structure of the delafossite-like $AFeO_2$ systems occur along the $b_m$ axis. Therefore, the nearest-neighbor (*nn*) exchange interactions ($J_1$) in the basal plane (Fig. 1b) play an essential role in the mechanisms responsible for lifting the magnetic frustration. In particular, we can speculate that the discussed [13] difference in the character of the monoclinic lattice distortion in the two ferrites is closely related to the change in the strength of the *nn* interaction.

The magnetism of real triangular magnets is characterized by a competition between the intra- ($J_i$) and inter-plane ($J_0$) interactions, and the single-ion anisotropy ($D$), whose sign and value determine the orientation of the spin plane relative to the crystal axes. The experimental evidence that the $b_m$ axis elongates in $CuFeO_2$ [13] implies that direct AFM exchange ($J_{dir} > 0$) diminishes and FM superexchange ($J_{sup} < 0$) will dominate. This mutual cancelation leads to a weakening of the nearest interactions ($J_1 > 0$). As a result, the anisotropy along the hexagonal $c_h$ axis ($D < 0$) gains importance, leading to additional frustration. According to Monte Carlo calculations [40], when $D/|J_1| > \xi_{crit} = 0.317$, the collinear ↑↑↓↓ (4SL) structure is expected to be stable. At the same time, in the case of $AgFeO_2$, the $b_m$ axis contracts, thus stabilizing AFM interactions along this axis. With increasing AFM exchange, corresponding to decreasing $D/|J_1| < \xi_{crit}$, the 4SL structure eventually becomes unstable to noncollinear incommensurate phases [40]. In particular, as will be discussed the next section, the enhancement of the nearest-neighbor Fe-Fe interaction at high temperatures ($T > T_{N2}$) gives rise to multiple-sublattice states with maximum AFM coupling between nearest neighbors, such as a partially disordered (PD) 5 sublattice structure (5SL) with a (…-↑↓↑↓-disordered-…) spin sequence in the $b_m$ direction.

### 3. Mössbauer data

#### A. *Paramagnetic temperature range* ($T > T_{N1}$)

The $^{57}$Fe Mössbauer spectra of AgFeO$_2$ measured in paramagnetic temperature range $T > T_{N1}$ (Fig. 3a) consist of a single quadrupole doublet with narrow ($W = 0.25(1)$ mm/s) and symmetrical components, indicating that all iron ions occupy in the ferrite structure equivalent crystal sites. The isomer shift $\delta_{300K} = 0.37(1)$ mm/s and quadrupole splitting $\Delta_{300K} = 0.66(1)$ mm/s correspond to the high-spin ions Fe$^{3+}$ ($S = 5/2$) in an octahedral oxygen environment with a very strong electric field gradient (EFG) [41]. Our calculations of the main components $\{V_{ii}\}_{i=x,y,z}$ of the EFG tensor $V^{lat}$, using crystal data for the high-temperature rhombohedral [42] and low-temperature ($T < T_{N2}$) monoclinic [42] lattices of AgFeO$_2$, have shown that, in addition to monopole lattice contribution $V^{mon}$, large weight have both dipole contribution $V^{dip}$, arising from the induced electric dipole moments ($p_O$) of oxygen O$^{2-}$ ions, and electronic contribution $V^{el}$ related to overlapping of ($ns$, $np$, $3d$)$_{Fe}$ and ($2s$, $2p$)$_O$ orbitals [33, 34] (see *Appendix*):

$$V_{ii}^{tot} = (1 - \gamma_\infty)\{V_{ii}^{mon} + V_{ii}^{dip}\} + (1 - R)V_{ii}^{el} ,  \qquad (13)$$

where $\gamma_\infty = -9.1$ and $R = 0.32$ – are Sternheimer's antishielding and shielding factors [30].

The $p_O$ moments were calculated with a self-consistent iterative method. The oxygen dipole polarizability $\alpha_O$ has been estimated from the fit of the calculated principal EFG components $|V_{zz}| \geq |V_{xx}| \geq |V_{yy}|$ to the experimental value of quadrupole splitting $\Delta = eQV_{zz}/2(1 + \eta^2/3)^{1/2}$, where $\eta \equiv (V_{xx} - V_{yy})/V_{zz}$ is the parameter of asymmetry of the EFG. Calculated partial values for $V^{mon}$, $V^{dip}$ and $V^{el}$ contributions and their dependences as a function of the polarizability of oxygen ions ($\alpha_O$) in AgFeO$_2$ matrix are shown in Fig. 3b. The best agreement between the theoretical and experimental values of quadrupole splitting at 300 K (Fig. 3b) was found for the polarizability $\alpha_O \approx 0.83$ Å$^3$ (for nominal charges $Z_O = -2$, $Z_{Ag} = +1$, and $Z_{Fe} = +3$ and the quadrupole moment of the $^{57}$Fe nuclei of $Q = 0.15$ barns [43]). The obtained high value of $\alpha_O$ agrees well with the data for other oxides with the delafossite structure [18]. The calculations show that the principal axis O$_z$ of the EFG tensor $V^{lat}$ is directed along the $c_h$ axis of the hexagonal AgFeO$_2$ unit cell (Fig. 4). The calculations of the principal components $\{V_{ii}\}_{i=x',y',z'}$, using the crystal parameters for low-temperature ($T < T_{N2}$) monoclinic phase, have shown that the angle between the principle axis O$_{z'}$ of the EFG tensor $V_{z'z'}^{tot}$ and $a_m$ axis in the ($ac$)$_m$ plane of the monoclinic unit cell is ~8 – 9$^0$, which is very close to the $c_h$ direction of the hexagonal unit cell. Therefore, we can conclude that a symmetry lowering $R\bar{3}m \to C2/m$ does not lead to significant changes in the values of the components $\{V_{ii}\}_{i=X,Y,Z}$ and their relative orientations in AgFeO$_2$ lattice.

We performed detailed measurements of the spectra at temperatures $T > T_{N1}$, and did not find any visible anomalies in the $\delta(T)$ and $\Delta(T)$ dependences. The isomer shift $\delta(T)$ gradually increase in accordance with the Debye approximation for the second-order Doppler shift [41]. This shows that there are no any electronic and structural transitions in this temperature interval.

### B. Critical spin fluctuations near $T \approx T_{N1}$

Fig. 5 shows the $^{57}$Fe Mössbauer spectra of AgFeO$_2$ taken in the temperature range below $T_{N1} \approx 14$ K, including the point of the second magnetic phase transition, $T_{N2} \approx 9$ K. A Zeeman structure with broadened components is clearly observed, evidencing the existence of a continuous distribution of the hyperfine magnetic field, $H_{hf}$, at the $^{57}$Fe nuclei. In the first stage of the spectral analysis, we reconstructed the magnetic hyperfine field distribution, $p(H_{hf})$ (Fig. 5), assuming a linear correlation between the quadrupolar shift ($\varepsilon_Q$) of the Zeeman components and the value of $H_{hf}$ [37]. From the temperature-dependences of the mean field $<H_{hf}>$ and dispersion $D_{p(H)} = \{\sum p(H_{hf})(H_{hf} - <H_{hf}>)^2 \delta H_{hf}\}^{1/2}$ (Fig. 6) of the resulting distributions $p(H_{hf})$, we determined the temperature ($T^*$) at which the magnetic hyperfine structure of the spectra disappears completely. The resulting value, $T^* \approx 19$ K, appears to be somewhat higher than the Neel temperature, $T_{N1}$, determined in the magnetic and thermodynamic measurements (Fig. 2). This finding is related to the persistence, in a narrow temperature range $T_{N1} \leq T \leq T^*$, of short-range magnetic correlations between Fe$^{3+}$ ions, which is usually observed for quasi-layered systems with frustrated exchange interactions [44]. This interpretation is discussed in detail below.

Assuming that the hyperfine field, $H_{hf}$, is proportional to the magnetization ($M_{Fe}$) of the iron sublattice at all temperatures, we can compare the experimental $H_{hf}(T)$ dependence with the critical-point behavior of the iron magnetization $M_{Fe}(T)$ at $T \to T_N$. We approximated, in the region near $T_{N1}$, the temperature dependence of the most probable hyperfine field $H_{hf}^{(max)}$, corresponding to the maximal value of the distribution $p(H_{hf})$, by a power law [45] as follows:

$$M_{Fe}(T)/M_{Fe}(4.6) \approx H_{hf}^{(max)}(T)/H_{hf}(4.6) = B(1 - T/T_N)^\beta, \quad (14)$$

where $H_{hf(4.6K)} = 485$ kOe is the value of the hyperfine field at $T = 4.6$ K, $B$ is a reduction factor which depends only on the lattice symmetry and spin value, and $\beta$ is a critical exponent. A reasonably good fit (Fig. 7) to a power law was obtained in the range $3 \cdot 10^{-3} \leq \tau \leq 0.57$, where $\tau \equiv 1 - T/T_N$ is the reduced temperature. The fitting leads to $B = 1.17(4)$, $T_N = 14.0(5)$ K, and a critical exponent $\beta = 0.28(4)$. The value of the parameter $B$ proves to be close to the theoretical value $B^{th} = 1.22$ for an ideal 2$D$ Ising magnet, while the resulting value of the critical exponent $\beta$ is significantly larger than the theoretical value, $\beta^{th} = 0.125$ [45], for two-dimensional magnetic systems. According to [45, 46], the possible reason for such a discrepancy can be related to the fact that both these values ($B$ and $\beta$) were determined outside the appropriate critical region, $0 < \tau \leq \tau_{crit}$, usually defined by $\tau_{crit} < 10^{-2}$. This critical region depends on the particular magnetic system and therefore, has to be determined for each case. We performed a least-squares fit to a power law (see Eq. (14)) to describe the experimental dependence of $H_{hf}^{(max)}(T)/H_{hf\,(4.6)}$ for various temperature regions, defined by the maximum value, $\tau_{max}$, of the reduced temperature $\tau$ (by successively omitting data points). The resulting variation of $\beta$ as a function of $\tau_{max}$ is shown in the inset to Fig. 7. The asymptotic behavior of $\beta$ is observed and at the same time, the critical exponent $\beta^*$ remains constant (within the error range). Therefore, a power law may yield reliable results, as observed only below $t_{max} \approx 0.15$. For $\tau > \tau_{max}$. To allow a direct comparison with theory, the experimental value of the exponent $\beta$ has to be adjusted while taking into consideration the selected range of reduced temperatures [45, 46] as follows:

$$\beta^* = \beta + A\nu \cdot (\tau_{max})^{\nu}, \quad (15)$$

where $\beta^*$ is the effective exponent that is related to the universal $\beta$ by $(\lim\beta^*)_{\tau \to 0} = \beta$; $A$ is the correction-to-scaling amplitude that depends on the features of the system; and $\nu$ is the universal correction-to-scaling exponent [45, 46]. This correction for the 2D Ising model ($\nu = 1$, $A \approx 0.21$) is presented in Fig. 7 (inset) by the orange, open circles. The evaluated averaged value, $\beta^* = 0.34(2)$, (the orange dashed line in the inset to Fig. 7) proves to be very close, but is slightly smaller than the theoretical value for 3D Heisenberg ($\beta^{th} = 0.365$) magnets [45].

The origin of the high critical parameter $\beta^*$ in the quasi-2D system AgFeO$_2$ is far from trivial, due to competition between several interactions such as the magnetic coupling between layers and the strength of the crystal field (single-ion anisotropy), which can lead to a range of $\beta$ values in the range $0.20 \leq \beta \leq 0.36$. The ideal 2D ($J_{\parallel} = 0$) isotropic Heisenberg system ($S_{x,y,z} \neq 0$) cannot be ordered at non-zero temperatures [45]. However, any deviation from the isotropic 2D system, such as a small anisotropy ($D \neq 0$) or interlayer exchange ($J_{\parallel} \neq 0$), favors the occurrence of long-range order within the magnetic layers at $T \neq 0$ K. Whether the order parameter of a system of weakly coupled magnetic layers shows a 2D Ising ($\beta \approx 1/8$) or a 3D ($\beta \approx 1/3$) [45] critical behavior depends on the relative strength of the anisotropy energy ($g\mu_B H_A$) and the interlayer exchange coupling ($J_{\parallel}$). The asymptotic value of $\beta^* \approx 0.34$ obtained in our experiments indicates that AgFeO$_2$ shows quasi-three-dimensional critical behavior and, thus, the magnetic phase transition ($T_{N1}$) in this oxide is governed by the interlayer interaction rather than by anisotropy as follows: $|J_{\parallel}| >> |D|$. Note that this interpretation agrees with the analysis of the revised magnetic properties of the triangular CuFeO$_2$ lattice [46]. It was suggested [46] that in addition to exchange couplings within layers ($J_\perp$), the exchange couplings between layers ($J_\parallel \approx 0.3 J_\perp$) also play a significant role in forming the magnetic structure of delafossite-like magnets. One can assume that the change in the character of interlayer exchange interactions when $Cu^+$ ions are replaced with $Ag^+$ is a possible reason for the observed drastic difference in the magnetic behavior of these ferrites. Further experimental and theoretical study is still required to reach a deeper understanding of critical dynamics in these low-dimensional iron-based multiferroics.

To examine the magnetic behavior of the iron sublattice at the first magnetic phase transition, we carried out a series of Mössbauer measurements just above the $T_{N1}$ temperature, $T_{N1} < T < T^*$. According to the obtained data (Fig. 8a), in addition to the paramagnetic quadrupole doublet with broadened components, it is necessary to introduce an unresolved magnetic hyperfine structure to provide a good description of the experimental spectra just above the "macroscopic" $T_{N1}$ temperature. Because the specific heat data (Fig. 2b) show that magnetic correlations are present in the range $T \geq T_{N1}$, we fitted the experimental spectra in the interval $T_{N1} < T < T^*$ using a stochastic relaxation model [47]. In this formalism, the magnetic hyperfine interactions are described as interactions between the nuclear magnetic moment and a randomly varying hyperfine magnetic field $H_{hf}(t)$. The time-dependent Hamiltonian for this relaxation is

$$\hat{H}(t) = \hat{H}_Q - g_n \mu_n H_{hf}^{(0)} \sum_i I_i f_i(t), \tag{16}$$

where $\hat{H}_Q$ is the quadrupolar hyperfine Hamiltonian, $g_n$ is the gyromagnetic factor of the nuclear state, $\mu_n$ is the nuclear Bohr magneton, $I_i$ are the nuclear spin projection operators onto the EFG principal axes, and $f_i(t)$ is a random function of time. We assumed that the $H_{hf}(t) = H_{hf}^{(0)} f_i(t)$ field fluctuates between the three principal directions of the EFG tensor with appropriate values $f_i(t) = \{1, 0, -1\}$ (isotropic relaxation). The lineshapes of the spectra depend on the following two parameters: a correlation time, $\tau_c$ (or frequency of the spin fluctuations $\propto 1/\tau_c$), and $H_{hf}^{(0)}$, which is the saturated hyperfine field when $1/\tau_c \to 0$ [48].

The observed coexistence of magnetic and paramagnetic subspectra close to the $T_{N1}$ point was interpreted as follows: antiferromagnetic clusters are possibly created in $AgFeO_2$ layers by two-dimensional spin correlations in the paramagnetic region near the temperature $T_{N1}$. A similar behavior has been previously observed for many low-dimensional systems [49]. It was shown that well above $T_N$, each spin relaxes in the local, rapidly fluctuating exchange field produced by its uncorrelated neighbors at the rate $1/\tau_c$, typically as large as $\sim 10^{11} - 10^{12}$ s$^{-1}$, thus resulting in a paramagnetic hyperfine structure (singlet or quadrupole doublet). As $T_{N1}$ is approached, the correlation length $\xi$ increases, i.e., small clusters of spins with a short-range order are formed, and these act as a single unit for which the magnetization relaxes slowly, as in superparamagnetic particles [50]. Thus, at $T \to T_{N1}$, where the system orders, both quantities $\tau_c$ and $\xi$ increase, and the characteristic frequency of the spin fluctuations decreases (inset to Fig. 8b). Such a critical slowing down of $1/\tau_c$ is expected to be more important in low-dimensional systems [49]. At $T \to T_{N1}$, the correlation length diverges and long-range order sets in. Although the magnetization of the crystal as a whole remains zero, it is possible to find increasingly large regions in which there is a net magnetization. From the fits of the spectra, we extracted the paramagnetic fraction ($I_{par}$) at different temperatures. Fig. 8b clearly shows that above $T_{N1}$, this paramagnetic fraction begins to increase sharply at the expense of the magnetic sub-spectrum, reaching a steady value of 100 % at $T^* \approx 21$ K. This behavior is consistent with superparamagnetism or superferromagnetism [50], i.e. the formation of nanosized magnetic domains with the randomly flipping direction of the magnetization under the influence of temperature. The relative temperature range $(T^* - T_N)/T_N$ of the dynamic critical region over which a slowing of the spin correlations time ($\tau_c$) occurs should be influenced by the ordering dimensionality, which is defined by β (Eq. 14). To verify the validity of this assumption, it would be interesting to compare the Mössbauer data for $AgFeO_2$ with those of other local resonance methods, such as NMR or ESR, sensitive to spin dynamics in the iron sublattice.

### C. Magnetic "intermediate" temperature range ($T_{N2} < T < T_{N1}$)

According to recent neutron diffraction data [12], at $T_{N2} \leq T \leq T_{N1}$, $AgFeO_2$ has a collinear, sinusoidally modulated spin structure with the propagation wave vector $Q$ along the monoclinic $b_m$ axis. We assumed that the hyperfine field distribution $p(H_{hf})$ observed in this temperature range (Fig. 9) is related to incommensurate spin-density-waves (SDW), in which iron ions at different iron sites, $x_n$, along the SDW propagation carry different values of magnetic moments ($\mu_{Fe}$). Assuming that the magnetic hyperfine field $H_{hf}(x_n)$ at each particular iron position is parallel and proportional to the magnetic moment, $\mu_{Fe}$, the modulation of the hyperfine field can be defined in terms of a Fourier series as [51]

$$H_{hf}(x_n) = \sum_{l=0}^{N} h_{2l+1} \sin[(2l+1)qx_n], \qquad (17)$$

where $h_{2l+1}$ represents the amplitude of the $i$-th ($i = 2l + 1$) harmonic, $q$ is the wave number of the SDW, $x_n$ is the relative $n$-th position of the iron ions along the direction of SDW propagation (for commensurate SDW, $x_n = nb_m$, and $n$ denotes the number of iron atoms in the direction of the $b_m$ axis). The spectra were fitted as a superposition of Zeeman subspectra, with hyperfine field values according to Eq. 17 for discrete ($qx_n$) values in the range $0 \leq qx_n \leq \pi/2$ [51].

An analysis of the experimental Mössbauer spectra using only a fundamental harmonic, $h_1$ (sine-modulation), did not yield a good fit (the blue dashed line in Fig. 9a). Thus, the $^{57}$Fe Mössbauer spectra can be described in terms of spin-density-waves (SDW) but with the inclusion of many high-order harmonics, $h_{2l+1,(l\neq 0)}$. This observation indicates that the real magnetic structure of the AgFeO$_2$ ferrite in the intermediate temperature range, $T_{N2} \leq T \leq T_{N1}$, appears to be more complicated than the early, supposedly purely sinusoidally modulated SDW [12] (the blue dashed lines on the distributions $p_{SDW}(H_{hf})$ and modulations $(H_{hf})_z(qx)$ of the hyperfine magnetic field in Fig. 9b). We obtained a series of least-squares ($\chi^2$) fits of the spectra, with harmonic amplitudes ($h_{2l+1}$) as the variable parameters. The number of Fourier components ($z$) in the fitting was increased until the experimental lineshape was satisfactorily reproduced. The good fits, shown in Fig. 9a, were obtained with six ($0 \leq l \leq 6$) Fourier components. The introduction of more components, i.e., $l \geq 7$, did not result in a significant decrease of $\chi^2$.

The observed slight asymmetry of the spectra (Fig. 9a) has been described assuming a linear correlation of the isomer shift, $\delta = \delta_0 + (\partial\delta/\partial H_{hf})H_{hf}$, and quadrupole shift, $\varepsilon_Q = \varepsilon_0 + (\partial\varepsilon/\partial H_{hf})H_{hf}$ with the hyperfine field, $H_{hf}$ [37]. Such correlations may be caused by an incommensurate modulation induced by magnetoelastic coupling with the SDW. A similar conclusion has been drawn from synchrotron X-ray diffraction experiments on a CuFeO$_2$ crystal, in which cooperative displacements of oxygen ions along the $b_m$ axis are induced in the partially distorted (PD) phase, as a result of magnetostriction [52]. It is interesting that the $^{57}$Fe Mössbauer spectra of the CuFeO$_2$ ferrite with the PD magnetic structure have very a similar profile to those of AgFeO$_2$ recorded at "intermediate" temperatures (Fig. 9a). Using the experimental values of $\{\delta_0, (\partial\delta/\partial H_{hf})\}_T$, $\{\varepsilon_0, (\partial\varepsilon/\partial H_{hf})\}_T$ and the average hyperfine field $<H_{hf}(T)>_{SDW}$, we estimated the mean values of $<\delta(T)>_{SDW}$ and $<\varepsilon_Q(T)>_{SDW}$ for each spectrum in the interval $T_{N2} \leq T \leq T_{N1}$. The $<\delta_{(10-18K)}>_{SDW} \approx$ 0.53 mm/s values thus obtained show good agreement with the data, $\delta_{30K} = 0.43(1)$ mm/s, for temperatures above $T_{N1}$ (accounting for the second-order Doppler shift). In contrast, the $<\varepsilon_Q(T)>_{SDW} \approx 0.03$ mm/s values appear to be significantly smaller than the quadrupole shift $\varepsilon_Q(=\Delta/2) \approx 0.16$ mm/s for $T > T_{N1}$. Taking into account that according to our calculations, monoclinic distortion of the AgFeO$_2$ lattice does not lead to significant changes in the values of the EFG components, the observed reduction of the $<\varepsilon_Q(T)>_{SDW}$ can be related to its angular dependence on the relative orientation of the principal axis, OZ, of the EFG tensor and $H_{hf}$. If $H_{hf} >> eQV_{zz}$ and the first-order quadrupole shift $\varepsilon_Q \equiv <3/2,m_I|\hat{H}_Q|3/2,m_I>$, the energy level $|3/2,m>$ is given by [53]

$$\varepsilon_Q = (-1)^{|m_I|+1/2}(\tfrac{1}{8}eQV_{zz}^{par})[3\cos^2\vartheta - 1 + \eta\sin^2\vartheta\cos 2\varphi], \qquad (18)$$

where $m_I$ are magnetic quantum numbers; $eQV_{zz}^{par}$ is the quadrupole splitting constant, which equals that in the paramagnetic state ($T > T_N$) if there is no distortion of the crystal lattice at $T_N$, and shows temperature-independent behavior below and above $T_N$; $\vartheta$ and $\varphi$ are the polar and azimuthal angles of the hyperfine field, $H_{hf}$, in the principal axes of the EFG tensor. Because the $Fe^{3+}$ ions occupy sites with a nearly axial symmetrical EFG tensor ($V_{xx} \approx V_{yy}$) in the $AgFeO_2$ structure, the parameter of asymmetry, η, was taken to be zero. Thus, we assumed that the spectral shape does not depend on the azimuthal angle φ. Taking the experimental values of $<\varepsilon_Q(T)>_{SDW}$, we calculated the average values of polar $\vartheta(T)$ angles for each temperature in the interval $T_{N2} \leq T \leq T_{N1}$. Figure 10 shows clearly that the obtained $\vartheta(T)$ values are well correlated with the $\phi(T)$ angles corresponding to the relative orientation of the SDW magnetization and the $c_h \| OZ$ axis deduced from ND data for $AgFeO_2$ [12].

Using the results of the above fitting, we calculated the distribution $p_{SDW}$. Fig. 9b shows a comparison between the hyperfine field distribution $p_{SDW}(H_{hf})$ and a similar distribution $p(H_{hf})$ of hyperfine fields, $H_{hf}$. An important feature of both distributions is the presence of several humps (Fig. 9b), which may be related to different $Fe^{3+}$ magnetic sublattices or a domain structure. The appearance of higher harmonics can be qualitatively understood within the Hamiltonian (1), which includes the main contributions of competing exchange interactions for the nearest neighbors, $J_{nn}$ ($\equiv J_1$), and next-nearest neighbors, $J_{nnn}$ ($\equiv J_{2,3}$), respectively (see Fig. 1b). We speculate that the observed profiles of SDW and the distribution $p_{SDW}(H_{hf})$ (Fig. 9b) can be understood using the domain wall or "soliton" model [54, 55]. Such an analysis has been performed in [54] for a simple three-dimensional Ising spin system with competing $J_{nn}$ and $J_{nnn}$ interactions, which exhibits modulated phases. It has been shown that when $|J_{nnn}/J_{nn}|$ is small, the ground state is non-distorted SDW, but when $|J_{nnn}/J_{nn}| \approx 0.5$, this state becomes marginally stable with respect to the domain-wall formation. The ground state is infinitely degenerate, corresponding to all possible ways to introduce domain walls. In general, the regularly spaced solitons build a soliton lattice with a soliton density that depends on the value of $|J_{nnn}/J_{nn}|$ [54, 55]. It must be noted that near $T_N$, the spin structure is almost sinusoidal, whereas, at lower temperatures, it contains a significant number of higher harmonics.

It should be noted that we cannot exclude another explanation for the observed $p(H_{hf})$ distributions with several maxima (Fig. 9b), which is based on the assumption of phase separation into magnetic and para- or superparamagnetic phases. Such an explanation was supposed earlier for the $CuFeO_2$ ferrite with the modulated, partially disordered (PD) magnetic structure in the range $T_{N2} < T < T_{N1}$ [56]. In inelastic neutron scattering experiments, a quasi-elastic magnetic scattering was observed in the PD phase, suggesting strong thermal spin fluctuations [56]. The same PD phase was recently found in the metallic triangular antiferromagnetic $Ag_2CrO_2$ using a muon-spin rotation and relaxation study ($\mu^+$SR) [45]. The spectra of this two-dimensional oxide were fitted by a combination of a cosine oscillation corresponding to a static but inhomogeneous internal field at the muon sites and two relaxing non-oscillatory signals for fluctuating $Cr^{3+}$ moments. This result reflects the dynamic character of the PD state, in which the distorted $Cr^{3+}$ spins are fluctuating too rapidly to be observed by $\mu^+$SR.

### D. *Low-temperature range, 4.6 K < T < $T_{N2}$*

The highly asymmetric profile of the experimental spectra at low temperatures below $T_{N2}$ (Fig. 5) reflects a high degree of correlation between the values of the magnetic hyperfine field and quadrupole shift ($\varepsilon_Q$) of the Zeeman components. Fig. 11 displays the temperature dependences of the correlation coefficients $\partial\delta/\partial H_{hf}$ and $\partial\varepsilon_Q/\partial H_{hf}$, obtained from the recovery of the distributions $p(H_{hf})$ (Fig. 5). The coefficient $\partial\delta/\partial H_{hf}$ is nearly equal to zero and does not depend on temperature, while the linear temperature-dependence $\partial\varepsilon_Q/\partial H_{hf} > 0$ shows a clear kink at $T \approx 8.5$ K, which almost completely coincides with the point $T_{N2} = 9$ K of the second magnetic phase transition [11,12]. According to [11], below $T_{N2}$, the magnetic moments of $Fe^{3+}$ ions form an elliptic cycloid with a period of ~500 Å, which propagates along the [010] direction in the hexagonal lattice. This change in the character of magnetic ordering within the iron sublattice of $AgFeO_2$ induces non-zero electric polarization [11].

The complex hyperfine magnetic structure for the $AgFeO_2$ ferrite significantly differs from the single magnetic sextet in the $CuFeO_2$ ferrite, which has the same delafossite-like structure and which is characterized by collinear magnetic ordering ↑↑↓↓ [10], excluding any spontaneous electric polarization [1,10]. At the same time, a similar hyperfine magnetic structure was observed for the perovskite-like ferrite $BiFeO_3$, possessing a non-collinear magnetic structure of the cycloid type [35, 36]. Thus, we can suppose that the observed inhomogeneous line broadenings of the Mössbauer spectra for $AgFeO_2$ reflect non-collinear spatially modulated spin ordering, which is one of the most important intrinsic features of "improper" multiferroics [4].

In the case of the $AgFeO_2$ ferrite, which has a crystal structure with the principal axis $O_z$ of the EFG tensor lying in the rotation plane of the iron magnetic moments [33, 34], the angle $\vartheta$ varies continuously in $0 \leq \vartheta \leq 2\pi$ interval (see Fig. 4), and the range of values of $\varepsilon_Q(\vartheta)$ would be the same for all lines in the Zeeman sub-spectrum, yielding homogeneous line broadenings. Therefore, to explain the sizeable observed spectral asymmetry (Fig. 5), we must take into account the angular dependence of the hyperfine magnetic field, $H_{hf}(\vartheta)$ [53]. Thus, the angular dependences of the parameters $\varepsilon_Q(\vartheta)$ and $H_{hf}(\vartheta)$ reflect the changes in the spatially modulated magnetic structure along the length of cycloid with respect to the hexagonal unit cell. For a model fitting of the Mössbauer spectra in the $T < T_{N2}$ range, we used a procedure similar to that applied earlier for the analysis of the $^{57}$Fe Mössbauer [33-37] and NMR [58] spectra of the multiferroic BiFeO$_3$, which also possesses a noncollinear magnetic structure of the cycloid type. We took into account the dependences of the quadrupole shift and hyperfine magnetic field on the polar angle $\vartheta$ between the direction of $H_{hf}$ (collinear, in the first approximation, with the direction of the Fe$^{3+}$ spins) and the principal axis O$_z$ of the EFG tensor, which coincides, according to our calculations, with the hexagonal axis of the AgFeO$_2$ lattice (Fig. 4) [53] as follows:

$$4\varepsilon_Q(\vartheta) = eQV_{zz}^{par}(3\cos^2\vartheta - 1)/2 + eQV_{zz}^{mag}, \tag{19a}$$

$$H_{hf}(\vartheta) = H_\parallel \cos^2\vartheta + H_\perp \sin^2\vartheta, \tag{19b}$$

where $V_{zz}^{par}$ is the principal component of the EFG tensor in the paramagnetic temperature range ($T > T_{N1}$); $H_\parallel$ and $H_\perp$ are the values of $H_{hf}$ oriented along and perpendicular to the principal axis O$_z$ (Fig. 12a). We have included in Eq. 19a an additional term, $eQV_{zz}^{mag}$, which can arise due to a local magnetoelastic coupling [59]. Because this additional term is caused by electron spins with a specific direction, it has to be axially symmetric, with the principal axis along the direction of magnetization of iron. Moreover, $\eta$ was set equal to zero for the sites with the nearly axially symmetric EFG tensor $V^{par}$ (see Eq. 18). Therefore, the spectral shape does not depend on the azimuthal angle $\varphi$. Given the very high value of the quadrupole splitting, we also used the second-order terms $a_\pm(H_{hf}, eQV_{zz}, \eta, \vartheta, \varphi)$ [36, 37].

Finally, we used Jacobian elliptic function (see Eq. 8) to describe the possible anharmonicity (bunching) of spatial distribution of Fe$^{3+}$ magnetic moments, which results from magnetic anisotropy in the ($zx$) plane of spin rotation (Fig. 12a). Using this procedure, the $^{57}$Fe Mössbauer spectra of the AgFeO$_2$ ferrite below $T_{N2}$ were fitted by the least-squares method (Fig. 13). It is possible to make a number of observations and comments based on the acquired data.

The best description of all these spectra was obtained by using sufficiently high values of the anharmonicity parameter $m_{4.6K} = 0.78(3)$ (Eq. 8), which remains almost constant in the entire temperature range (Fig. 14). To visualize the effect of the uniaxial anisotropy, which results in the distortion of the circular cycloid, on $H_{hf}$, the dependences $H_{hf}^{(z)} \propto \text{sn}[(\pm 4K(m)/\lambda)x, m]$ and $H_{hf}^{(x)} \propto -\text{cn}[(\pm 4K(m)/\lambda)x, m]$ in terms of the Jacobi elliptic functions for the easy-axis anisotropy ($D > 0$) are shown together with $\sin(qx)$ in Fig. 15. It is clearly observed that, in spite of the very small value of the single-ion anisotropy, the functions $H_{hf}^{(z)}$ and $H_{hf}^{(x)}$ themselves are significantly deviated from harmonic behavior. Depending on the parameter $m$, the spin modulation $H_{hf}^{(z)}$ ($H_{hf}^{(x)}$) changes from a pure cosine (sine) wave ($m = 0$; $D_z = 0$) to a square wave ($m \to 1$, $D_z \neq 0$) in which spins bunch along the $V_{zz}$ direction (Fig. 12b). Another manifestation of anisotropy is the dependence of the Δϑ(x) angles between neighboring iron spins along the cycloid propagation, presented in Fig. 12a. Without easy-axis anisotropy ($D = 0$), the Δϑ angle defining the relative orientation of spins for the given wave vector is constant, $\Delta\vartheta_0 = qa$. However, for the case of easy-axis anisotropy, the cycloid with $\vartheta_0$ no longer describes the minimal energy of spin arrangement, due to the admixture of higher-order harmonics. The anisotropy tends to align the spins along the easy axis (OZ), thereby distorting the originally perfect cycloid into one with modulated rotation angles $\Delta\vartheta(x) = \Delta\vartheta_0 + \xi(x)$ (where $\xi(x)$ depends on the value of $D/\langle J \rangle$). In Fig. 15, the non-uniform population density of spin angles $\rho(\vartheta) \propto 4K(m)/\lambda(1 - m\cos^2\vartheta)^{1/2}$ peaks at 90° and 270° because the iron spins prefer to align along $\pm O_z$ directions.

Anharmonicity is related to the constant $K_u = NS^2|D|/V$ of the uniaxial anisotropy as follows [23, 58]: $K_u = 16mAK^2(m)/\lambda^2$, where $K(0.78) = 2.21$, and $\lambda \approx 500$ Å [11]. Taking into account that $A \approx NS^2(b_m)^2\langle J_1\rangle/V$ with the monoclinic unit cell parameter $b_m = 3.03$ Å [11] we have estimated the value $D/J_1 \approx 16mK^2(m)(b_m)^2/\lambda^2 \approx 2.2\cdot10^{-3}$ that appears to be substantially smaller than the critical value $(D/|J_1|)_{crit} \approx 0.27$, below which the collinear ↑↑↓↓ 4SL phase of $CuFeO_2$ is not energetically stable and is replaced by more stable complex noncollinear (CNC) spin structures [60]. Below the second critical value of $(|D/J_1|)_{crit} \approx 0.08$, the cycloid-like structure has a lower energy than the CNC phase [60]. As an example, we can refer to the $CuFe_{0.965}Ga_{0.035}O_2$ system, in which $(|D/J_1|) \approx 0.04$ [6], exhibiting a helicoidal magnetic structure and multiferroic behavior. As discussed in [6], this reduction of the $(|D/J_1|)$ ratio must be one of the reasons for the disappearance of the collinear 4SL magnetic ground state. If the exchange stiffness constant to estimate as follows: $A \approx 3/2(k_BT_N/R_{Fe-Fe})$, we arrive at an estimation of the single-ion anisotropy parameter $D$ as $D \sim -0.017$ meV, that is twice as much as our theoretical prediction ($-0.008$ meV) for trigonally distorted $FeO_6$ octahedra in $AgFeO_2$ given nonpolarized oxygen ions. One of the most probable explanations both for this deviation and an unexpected spread of the anisotropy parameters in ferro-delafossites can be related with a markedly large and sensitive covalent contribution of the electrically polarized oxygen ions [18, 61] which seems to be a specific property of delafossite structure as compared with many other oxides. All the above estimations may be useful in explaining the possible reasons for the different character of magnetic ordering in the two delafossite-like $CuFeO_2$ and $AgFeO_2$ ferrites.

The approximated saturation value of the hyperfine field, $H_{hf(T \to 0)} \approx 484$ kOe, is anomalously low for high-spin ferric ions in octahedral oxygen coordination, for which $H_{hf}(0)$ is usually approximately 540-568 kOe (Fig. 16), corrected for covalence effects [62]. This ~10% spin reduction cannot be explained only by covalency effects and may be partially attributed to zero-point spin reduction, which has been predicted to be large in low-dimensional antiferromagnets [63]. However, our estimations of the zero-point spin fluctuations (not shown) point to a very small contribution to the reduction of the $H_{hf}(0)$ value for $AgFeO_2$.

To discuss other possible reasons for the observed reduction of the $H_{hf}(0)$ value, we took into account the fact that the $\boldsymbol{H}_{hf}$ hyperfine field is the vector sum of several contributions [62, 64]:

$$\boldsymbol{H}_{hf} = \boldsymbol{H}_F + \boldsymbol{H}_{cov} + \boldsymbol{H}_{STHF}, \qquad (20)$$

where $\boldsymbol{H}_F$ is the free-ion field produced by the Fermi contact interaction, the covalent contribution, $\boldsymbol{H}_{cov}$, arises from the difference in overlap and transfer effects of the spin-up and down s-orbitals of iron. These two contributions are proportional to the vector $<S>$ directed along the thermally averaged 3d spins. $\boldsymbol{H}_{STHF}$ is the supertransferred contribution resulting from all single-bridged nearest ferric neighbors "n", each proportional to the electronic spin $<S_n>$, on the neighboring site:

$$H_{STHF} = \sum_n B_n (\langle S_n / S \rangle), \qquad (21a)$$

$$B_n = \left\{ (H_\sigma^{(n)} - H_\pi^{(n)}) \cos^2 \theta_n + H_\pi^{(n)} \right\} + H_{sd}^{(n)} = z_n (h_{sthf}^{(n)} + h_{dir}^{(n)}), \qquad (21b)$$

where $B_n$ is a positive scalar field parameter depending on the "superexchange" iron-ligand-iron bond angle, $\theta_n$, and "direct" iron-iron bond distance; $H_\sigma$ and $H_\pi$ parameters arise from the spin-polarization of iron s-orbitals, caused by the ligand p-orbitals that have been unpaired by spin transfer, via σ and π bonds, into unoccupied 3d orbitals on the neighboring cations; $H_{sd}$ ($h_{dir}$) is the direct contribution arising from the overlap distortions of iron s-orbitals by 3d-orbitals of the neighboring cation; and $z_n$ is the number of nearest neighbors.

The existence of the angular dependence of the $H_{STHF}$ field has been shown experimentally in several studies of rare earth (R) orthoferrites $RFeO_3$ [64, 65]. In these perovskite-like compounds, there is no direct overlap of iron orbitals, so one can neglect the direct Fe-Fe hyperfine field ($h_{dir} \approx 0$) and consider only supertransferred contributions, $h_{sthf}$, from six equivalent Fe-O-Fe bonds ($z = 6$). In the magnetic temperature range ($T < T_N$), each iron ion is surrounded by six $Fe^{3+}$ ions with the same direction of collinear magnetic moments (see the upper part of Fig. 16). According to the results of *Boekema et al.* [64] on the magnetic hyperfine field at $^{57}Fe$ sites of $LaFe_{1-x}Ga_xO_3$, the average of the supertransferred contribution, $h_{sthf}$, per iron ion is equal to 9.1 kOe. In addition, the calculations of *Moskvin et al.* [66] for the $RFeO_3$ ferrites have shown that $H_\sigma = 60.2$ kOe and $H_\pi = 9.8$ kOe. Thus, $h_{sthf} = 8.8$ kOe for $LaFeO_3$ ($\cos^2\theta = 0.846$), which is in accordance with the work [64]. Using the average value of $h_{sthf} = 8.95$ kOe, we evaluated ($H_F + H_{cov}$) = 564 - 6×$h_{sthf}$ = 494 kOe. This value, corresponding to the hyperfine field at $^{57}Fe$ sites in the "free" ($FeO_6$) octahedra, is used as a reference value (Fig. 16) in our subsequent calculations.

In the delafossite-like structure of the $CuFeO_2$ ferrite, each iron cation is surrounded by six nearest $Fe^{3+}$ neighbors lying in the hexagonal ($a_h b_h$) plane (Fig.1). Below $T_{N2}$, $CuFeO_2$ demonstrates a collinear four-sublattice magnetic structure ↑↑↓↓, where two of the six nearest $Fe^{3+}$ cations have the same spin direction as central iron cation [40]. The remaining four $Fe^{3+}$ neighbors have the opposite spin direction (the middle part of Fig. 16). As a result, in the immediate surrounding of the central iron cation, there are two pairs of mutually compensated $Fe^{3+}$ spins that give a resultant zeroth contribution to the experimental value of the hyperfine field $H_{hf} \approx 515$ kOe. In this case, the difference $\Delta H_{exp}\{\equiv H_{hf} - (H_F + H_{cov}) = 515 - 494\} = 21$ kOe corresponds to the positive partial contribution of two $Fe^{3+}$ neighbors (Fig. 16). Taking into account the fact that the $\theta$ angle for the Fe-O-Fe bonds is approximately $90^0$, we can neglect the supertransferred contributions $h_{sthf}$ (according to Ref. [66]. The tiny positive field, $H_\pi$, can be compensated by the strong negative contribution of potential s-d exchange at $\theta \to 90^0$). Therefore, the experimental difference $\Delta H_{exp}$ is equal to the direct contributions of the two nearest iron neighbors as follows: $2h_{dir} = \Delta H_{exp} + \Delta H(0) \approx 24$ kOe, and $h_{dir} \approx 12$ kOe.

$AgFeO_2$ exhibits noncollinear magnetic order (see lower part of Fig. 16), in which among six nearest iron neighbors of the central iron cation, there are two pairs with compensated spins, giving a total contribution of zero to the $H_{hf}$ field. As a result, the $H_{hf}$ value for the $AgFeO_2$ ferrite can be presented as $H_{hf} \approx (H_F + H_{cov}) - 2h_{dir}\cos\xi$, where the angular part $\sim \cos\xi$ is a measure of deviation from the collinear structure, $\xi = 2\pi q$ (Fig. 16). Taking into account that for the $AgFeO_2$ magnetic structure below $T_{N2}$ the incommensurate propagation wave vector $q = 0.2026$, we obtain $\xi \approx 74.25^0$. Substituting $(H_F + H_{cov}) = 494$ kOe, $h_{dir} = 12$ kOe, and $\cos\xi = 0.271$, we obtain $H_{hf} = 488$ kOe, which is in very good agreement with our experiment $H_{hf}(T \to 0) \approx 484$ kOe (slight difference of these values may be associated with zero-point spin reduction $\Delta H(0)$, indicated by orange line in Fig. 16). This result suggests that the observed difference between the $H_{hf}$ values for $AgFeO_2$, $CuFeO_2$, and $RFeO_3$ ferrites is mainly related to their local magnetic structure, whereas the magnetic dimensionality plays a minor role.

### E. *Origin of anisotropic hyperfine magnetic fields in AgFeO₂*

According to our calculations, $V_{zz} > 0$, that is the maximum value of quadrupole shift $\varepsilon_Q(\vartheta)$ is attained at $\vartheta = 0^0$ (Eq. 19 a). Taking into account the positive value of the correlation coefficient $\partial\varepsilon_Q/\partial H_{hf} \approx +1.6\times10^{-4}$ mm/s/kOe (see Fig.11), one may conclude that maximum hyperfine field $H_{hf}$ is attained for the spins of $Fe^{3+}$ ions that are directed along the $V_{zz}$ ($H_\parallel$) axis, i. e. $H_\parallel > H_\perp$, similar to the elliptical polarization (Fig 12c). Note that a similar character of hyperfine field anisotropy ($H_\parallel > H_\perp$) is observed in NMR spectra for $BiFeO_3$ ferrite having a similar noncollinear magnetic structure of the cycloid type [35, 36].

The analysis of the distribution $p(H_{hf})$ (Fig. 13) shows the strong anisotropy of hyperfine field $H_\parallel = 499(1)$ kOe and $H_\perp = 476(1)$ kOe (at 4.7 K), in addition, the difference $\Delta H_{hf} = (H_\parallel - H_\perp)$ increases with the temperature (Fig.14). The magnetic ordering of iron sublattice does not induce a significant additional "magnetic" contribution to the EFG ($eQV_{zz}^{mag}$ = 0.08(9) mm/s). At the same time, the contribution $eQV_{zz}^{par}$ = 1.24(2) mm/s related to the symmetry of lattice proves to be very close to the average value $eQV_{zz}$ = 1.31(1) mm/s obtained in paramagnetic temperature range above $T_{N1}$. Therefore, the transition to the magnetically-ordered region $T < T_{N1}$ does not cause a significant distortion in the local environment of the iron ions.

Assuming that a hyperfine coupling tensor, $\tilde{A}$, specifying the coupling between the nuclear spin and the electronic spin ($S_{Fe}$), is diagonal with respect to the principal axes of the EFG tensor with axial symmetry, the hyperfine field $\boldsymbol{H}_{hf}$ can be written as follows [59, 67],

$$\boldsymbol{H}_{hf} = \tilde{A} \cdot \boldsymbol{S}_{Fe} = \boldsymbol{i}A_{xx}S_x + \boldsymbol{j}A_{yy}S_y + \boldsymbol{k}A_{zz}S_z = A_\perp(\boldsymbol{i}S_x + \boldsymbol{j}S_y) + A_\parallel \boldsymbol{k}S_z, \qquad (22)$$

where $S_{x,y,z}$ are projections of the Fe spin moment on the principal axes of the EFG tensor; $A_{xx} = A_{yy} \equiv A_\perp$ and $A_{zz} \equiv A_\parallel$ are the values of the components of the hyperfine coupling tensor $\tilde{A}$ corresponding to the orientation parallel and perpendicular to the principal axis $O_z$ of the EFG tensor. The expression (14) clearly suggests that the observed anisotropy of the hyperfine field can be analyzed in terms of anisotropy of the hyperfine coupling tensor $\tilde{A}$ and anisotropy of iron magnetic moment. In this formalism it is assumed $S_x = S_{Fe}\sin\theta\cdot\cos\varphi$, $S_y = S_{Fe}\sin\theta\cdot\sin\varphi$ and $S_z = S_{Fe}\cos\theta$, where $\theta$ and $\varphi$ are the polar and azimuthal angles of the $\boldsymbol{S}_{Fe}$ direction with respect to the principal EFG axes. Therefore, if $A_\perp \neq A_\parallel$, the resulting hyperfine field is no longer parallel to the direction of the iron magnetic moments except in the case of alignment along the $c_h$ axis, or in the $(ab)_h$ plane. In other cases, the angle $\vartheta$ which the $H_{hf}$ field makes with the principal $O_z$ axis (see Fig. 4) can be written as $\tan\vartheta = (A_\perp/A_\parallel)\tan\theta$, and the magnitude of the hyperfine field will be given by $H_{hf} = S_{Fe}(A_\parallel^2\cos^2\theta + A_\perp^2\sin^2\theta)^{1/2}$. However, taking into account that $H_\perp/H_\parallel \approx 1$, in first order, only the component $\tilde{A}\cdot\boldsymbol{S}_{Fe}$ parallel to the spin direction will affect the magnitude of the $H_{hf}$ field, we can rewrite the expression for $H_{hf}(\|\boldsymbol{S}_{Fe}) = A_\parallel S_{Fe}\cos^2\theta + A_\perp S_{Fe}\sin^2\theta$, that is similar to the Eq. 19b ($\theta \approx \vartheta$).

Both principal $A_\perp$ and $A_\parallel$ components include isotropic ($A_\parallel^{iso} = A_\perp^{iso}$) term, related to the Fermi contact hyperfine field due to the polarization of the inner $ns$ electrons by exchange interaction with the unpaired spins of the $d$ electrons, and anisotropic ($A_\parallel^{anis} \neq A_\perp^{anis}$) terms. Usually the anisotropy of the hyperfine tensor $\tilde{A}$ for the $Fe^{3+}$ ions having a spherically symmetric $d^5$ configuration is related to a distant dipole contribution $A_{ii}^{dip} = \mu_{iFe}D_{ii}$, where $D_{ii}$ ($i = z, x$) are the principal components of the lattice sums and $\mu_{iFe}$ are the projections of the magnetic moment of iron onto corresponding principal axes of the EFG tensor. Using the structural data for $AgFeO_2$ [13, 37] we arrive at $D_{zz} = -0.223$ Å$^{-3}$ and $D_{xx} = 0.112$ Å$^{-3}$ that corresponds to maximal value of $H_{an} = 5$ kOe that appears to be distinctly smaller than the experimental values ~30 kOe. Therefore, the observed anisotropy of hyperfine coupling in $AgFeO_2$ ferrite cannot be explained using only the dipole contribution $H_{dip}$.

The local anisotropic $\tilde{A}^{anis}$ coupling tensor is usually expressed as a superposition of spin and orbital terms:

$$\tilde{A}^{anis} = \tilde{A}^{spin} + \tilde{A}^{orb}, \qquad (23)$$

where for one $d$-electron

$$A_{ij}^{spin} = \frac{2g_s\mu_B}{21}\left\langle\frac{1}{r^3}\right\rangle\langle 3l_il_j - l^2\delta_{ij}\rangle, \qquad (24a)$$

$$A_{ij}^{orb} = 2\mu_B\left\langle\frac{1}{r^3}\right\rangle\langle \tilde{g}_{ij} - g_s\delta_{ij}\rangle, \qquad (24b)$$

where $\langle 1/r^3\rangle$ is the average value of $r^{-3}$ for the 3$d$ orbitals, $g_s \approx 2$ is the spin $g$-factor, $\mu_B$ the Bohr magneton, $g_{ij}$ the $\tilde{g}$-tensor. The first term is produced by the dipolar interaction of the $d$-electron spins with the nucleus, and does not vanish only when the $d$ orbitals are such that the spin density is aspherical. This term is related to the electronic part of the EFG, $V^{el}$, which arises from an aspherical charge density, and usually represented as $\tilde{A}^{spin} \approx -g_s\mu_B\cdot(V^{el})_{3d}$ [68]. For isolated high spin $Fe^{3+}$ ions having spherical 3$d$ electron distribution or for ideal $FeO_6$ octahedra with orbital singlet ground 3$d$-state $^6A_{1g}$ the $\tilde{A}^{spin}$ term turns into zero. However, specific feature of the high spin $3d^5$ configuration for $Fe^{3+}$ ions is that these have the only $^6A_{1g}$ term with maximal spin $S_{Fe} = 5/2$. It means that the intra-configurational low-symmetry crystal field for $^6A_{1g}$ term due to more distant monopole, dipole,… contributions cannot induce orbital anisotropy and nonzero $\tilde{A}^{spin}$, though these effects can be of a principal importance for the EFG. However, covalent effects due to the anion-cation $Fe^{3+}$-$O^{2-} \rightarrow Fe^{2+}$-$O^-(\underline{L})$ charge transfer and overlap in the low-symmetry distorted $FeO_6$ octahedra produce inter-configurational mixing effects, in particular, mixing of the $(3d^5)^6A_{1g}$ term with the orbitally active $(d^6\underline{L})$ $^6T_{1g}, ^6T_{2g}$ terms for the charge transfer configuration $d^6\underline{L}$, where $\underline{L}$ denotes the oxygen hole. As a result, we arrive at nonzero $A^{spin}$.

The orbital term $\tilde{A}^{orb}$ corresponds to the magnetic field produced at the nucleus due to orbital currents. In general, the low-symmetry crystal field quenches the orbital angular moment ($<L> = 0$), however, for orbitally nondegenerate states the spin-orbital coupling partially "restores" the orbital contribution into effective magnetic moment $\mu = \mu_B\tilde{g}S$ with effective $\tilde{g}$-tensor instead of $g_S \approx 2$. Thus, the anisotropic orbital term in $H_{hf}$ arises from the anisotropy of the $\tilde{g}$-tensor. The orbital contribution to the magnetic field on the Fe nucleus can amount rather large values of $H^{orb} \approx 20$ kOe given $\Delta g = 0.01$ [68]. However, usually the anisotropy of the $\tilde{g}$-tensor is rather small for so called S-type ions ($Fe^{3+}$, $Mn^{2+/4+}$, $Cr^{3+}$, $Ni^{2+}$) with $A_{1g}$ orbital singlet ground state. It is worth noting a close relation of the $\tilde{g}$-tensor anisotropy with the single-ion anisotropy. For axially symmetric $Fe^{3+}$ centers we have

$$g_\parallel - g_\perp = -\frac{2D}{\xi_{3d}}, \quad (25)$$

where $\xi_{3d} \approx 60$ meV is the one-electron spin-orbital coupling parameter. For the easy axis type single ion anisotropy $D < 0$ and we arrive at $g_\parallel - g_\perp > 0$ and $A_\parallel^{orb} - A_\perp^{orb} > 0$.

Unfortunately, we do not have any information concerning the value of the $\tilde{g}$-tensor anisotropy for the $AgFeO_2$. The ESR study of $CuFeO_2$ ferrite [69] with the same as $AgFeO_2$ delafossite-like structure, has revealed an almost isotropic g-factor ($g \approx 2$) at low temperatures, that seems to rule out the presence of significant orbital contribution to the $A^{anis}$ coupling constant, though the non-zero value of single-ion anisotropy $D \approx -(0.02 - 0.03)$ meV [69] implies small non-zero anisotropy $g_\parallel - g_\perp \approx 0.001$ and $H^{orb} \approx 2$ kOe. Thus, the main contribution to the anisotropy of the hyperfine coupling tensor should be related with a local intra-cluster $FeO_6$ spin-dipole term whose magnitude first depends on the cation-anion covalence effects. Similarly to single-ion anisotropy for weakly distorted $FeO_6$ octahedra with nonpolarized ligand one may introduce a simple linear parameterization for the $\tilde{A}^{spin}$ tensor in the principal axes system ($O_{x,y,z} \parallel C_4$) as follows:

$$A_{ii}^{spin} = a_E \epsilon_{ii}, \quad A_{ij}^{spin} = a_{T_2} \epsilon_{ij}, \quad (26)$$

where $\epsilon_{ij}$ are the components of the octahedron deformation tensor (*see* Introduction). For $(FeO_6)$ octahedra with equal cation-ligand separation as in $AgFeO_2$ and $CuFeO_2$ one may use another parametrization approach, based on the so-called superposition model, as follows:

$$A_{ij}^{spin} = \frac{1}{2} a \sum_{n=1}^{6} \left(3 x_{in} x_{jn} - \delta_{ij}\right), \quad (27)$$

where $x_{in}$, $x_{jn}$ are the Cartesian coordinates of the cation – $n$-th ligand bond. For trigonally distorted octahedron in the coordinate system with $O_z \parallel C_3$ we arrive at a simple relation:

$$A_{zz}^{spin} = 3a\left(3\cos^2\theta_n - 1\right), \quad (28)$$

where $\theta_n$ is the polar angle of the cation-ligand bond. It is worth noting that the contribution turns into zero for the critical angle $\theta_n = \theta_{crit} = \cos^{-1}(1/\sqrt{3}) \approx 55^0$. However, the above parametrization does not work for $FeO_6$ octahedra with polarized ligands characterized by nonzero electric dipole moment due to a shift of the valent electron shell, or a local *s-p* hybridization. In such a case one may use the superposition model for the dipole-induced EFG [33, 34] considering that $A_{ij}^{spin} \propto V_{ij}^{dip}$. Assuming that all the six ligand dipoles in the trigonally distorted $FeO_6$ octahedra are oriented parallel $O_z \parallel C_3$ with an "antiferromagnetic" ordering of the three "upper" and three "bottom" dipoles, respectively, we arrive at a rather simple relation

$$A_{zz}^{\text{spin}} = b \sum_{n=1}^{6} \frac{(3\cos\theta_n + 5\cos 3\theta_n)}{8} d_z(n), \quad (29)$$

where $d_z(n) = \pm 1$ for positive and negative orientation of the dipole, respectively. To demonstrate the workability of this simple two-parameter model we will assume its applicability both to $AgFeO_2$ and $BiFeO_3$ considering the geometrical factor to be the only cause for different anisotropy of the local fields: $\Delta H_{\text{anis}} = H_\parallel - H_\perp = 30$ kOe ($AgFeO_2$), 5 kOe ($BiFeO_3$ [35, 36]); $\theta_n = 59.5^0$ ($AgFeO_2$), $45.5^0$ ($BiFeO_3$). Experimental data are nicely explained, if $a = 1.8$ kOe, $b = 3.6$ kOe. We see that our conjecture points to a leading contribution of the dipole-induced term. Furthermore, it does uncover the origin of a strong difference in $\Delta H_{\text{anis}}$ for $AgFeO_2$ and $BiFeO_3$. In the delafossite $\theta_n > \theta_{\text{crit}}$ and the both terms work with the same sign, while in $BiFeO_3$ $\theta_n < \theta_{\text{crit}}$ and we arrive at a partial compensation of the two contributions.

For further investigation both of the origin and magnitude of the anisotropic magnetic interactions in $3R$-$AgFeO_2$ the measurements that can directly probe the electronic states of $Fe^{3+}$ and $O^{2-}$ ions, such as x-ray-absorption and x-ray photoemission spectroscopy, optical spectroscopy, and $^{17}O$ NMR study are required.

## CONCLUSIONS

In summary, we have carried out detailed $^{57}Fe$ Mössbauer measurements on polycrystalline samples of $3R$-$AgFeO_2$ that allowed us to elucidate different unconventional features of the electronic and magnetic structure, as well as spin and hyperfine interactions in this delafossite as compared with its analogue $CuFeO_2$.

The ferrite exhibits very strong magnetic frustration ($\Theta_{CW}/T_N$), which can be related to the strengthening of the AFM contribution in the nearest-neighbor Fe-Fe exchange interactions ($J_1^{(1)}$) in the basal plane. The asymptotic value, $\beta^* \approx 0.34$, for the critical exponent obtained from Mössbauer measurements indicates that $3R$-$AgFeO_2$ shows quasi-three-dimensional critical behavior. Mössbauer measurements just above the $T_{N1}$ temperature $T_{N1} < T < T^* \approx 19$ K show unresolved magnetic hyperfine structure, indicating the occurrence of short-range magnetic correlations; small clusters of iron spins are created, for which the magnetization relaxes slowly, as in superparamagnetic particles. In the intermediate temperature range of $T_{N2} < T < T_{N1}$, the $^{57}Fe$ Mössbauer spectra can be described in terms of SDW but with the inclusion of many high-order harmonics, indicating that the real magnetic structure of the ferrite in this intermediate temperature range appears to be more complicated than the previously supposed purely sinusoidally modulated SDW.

The line broadenings and spectral asymmetry at $T < T_{N2}$ arise from the spatial modulation of the electric hyperfine interactions and the intrinsic anisotropy of the magnetic hyperfine field at the $^{57}Fe^{3+}$ sites along the cycloid propagation vector. The model fitting shows a high degree of the anharmonicity in the cycloid ($m \approx 0.78$), which is related to rather large magneto-crystalline single-ion anisotropy. Comparison of the theoretical estimations within conventional models that imply nonpolarized oxygen ions and different experimental data points to a markedly large and sensitive covalent contribution of the electrically polarized oxygen ions which seems to be a specific property of delafossite structure as compared with many other oxides. We argue that such oxygen dipoles calculated self-consistently provide a large and decisive contribution to the EFG tensor for Fe nuclei while conventional monopole (or point charge) and electronic contributions partially compensate each other.

The hyperfine field, $H_{hf}$, reveals a puzzlingly large anisotropy with $\Delta H_{hf} = (H_\parallel - H_\perp) \approx 30$ kOe that cannot be related to only the dipole contribution from the magnetic neighbors or conventional orbital contribution $\propto g_\parallel - g_\perp$. This anisotropy is shown to be related with a local intra-cluster $FeO_6$ spin-dipole term that implies a conventional contribution of the cation-anion covalence effects induced by nonpolarized oxygen ions and even larger contribution of the polarized oxygen. We propose a simple two-parametric formula to describe the dependence of $\Delta H_{hf}$ on the distortions of the $FeO_6$ octahedra. A variety of different results evidencing a specific role of the oxygen polarization in ferro-delafossites seems to be one of the main outcomes of the paper. The very large difference between the $H_{hf}$ values observed for $AgFeO_2$ and $CuFeO_2$ ferrites is mainly related to their local magnetic structure, whereas the magnetic dimensionality plays a minor role.

Our study shows once more that Mössbauer spectroscopy is not only an useful tool to probe the local crystal structure and magnetic interactions of the iron cations in delafossites but a technique that is capable to uncover novel unexpected effects.

## ACKNOWLEDGMENTS

This work was supported by the Russian Foundation for Basic Research, grant # 14-03-00768. One of the authors (ASM) acknowledges the support by the Government of the Russian Federation, Program 02.A03.21.0006 and by the Ministry of Education and Science of the Russian Federation, projects nos. 1437 and 2725.

## *APPENDIX*

The lattice contribution to the EFG at the $Fe^{3+}$ sites was calculated using a monopole-point-dipole model [33, 34]. The monopole contribution ($V_{ij}^{mon}$) is given by

$$V_{ij}^{mon} = \sum_k Z_k (3x_{ik}x_{jk} - \delta_{ij}r_k^2)/r_k^5, \tag{1A}$$

where $Z_k$ is the charge and $x_{ik}$, $x_{jk}$ are the Cartesian coordinates of the $k$-th ion with a distance $r_k$ from the origin located at a given site, $\delta_{ij}$ is the Kronecher index. The dipole contribution $V_{ij}^{dip}$ is

$$V_{ij}^{dip} = \sum_k -3[(x_{ik}p_{ik})(5x_{ik}x_{jk} - \delta_{ij}r_k^2)/r_k^7 - (x_{ik}p_{ik} + x_{jk}p_{jk})/r_k^5], \tag{2A}$$

where $p_{ik}$ is the $i$-th component of the induced dipole moment on the $k$-th ion and the other symbols have the same meaning as in *Eq.* (1A). The components of the induced dipole moment are equal to

$$p_{ik} = \sum_i \alpha_{ij}^k E_j^k, \tag{3A}$$

where $\alpha^k$ is the polarizability tensor of the $k$-th ion and $E_j^k$ is the $j$-th component of the total electric field at the $k$-th ion. Since the induced dipole moments contribute to the electric field themselves, they have been calculated with a self-consistent iterative process. Due to the local symmetry at the sites of the $Fe^{3+}$ and $Ag^+$ cations in the 3$R$-AgFeO$_2$ lattice, an electric field $E^k$ exists only at the oxygen sites. Thus, only the oxygen ions contribute to $V_{ij}^{dip}$.

The $\alpha^k$ value is not well known and was estimated from the best fit of the theoretical EFG's to the measured data. In our calculations we used values of $\alpha_O$ in the range 0.1 - 1.0 Å$^3$. The lattice sums (1A-2A) were calculated with the spherical boundary method in which the summation is carried out by considering the contributions from all lattice sites inside a sphere given radius ($r$). The calculated contributions to the EFG were diagonalized and the resulting principal values of $V_{ii}^{lat}(= V_{ii}^{mon} + V_{ii}^{dip})$ were designated according to the usual convention $|V_{zz}| \geq |V_{xx}| \geq |V_{yy}|$. We assumed that the all ions in the AgFeO$_2$ ferrite have charges equal to their formal oxidation states.

This result indicates the need to consider local valence contribution $V_{zz}^{el}$ to the total EFG due to the overlap of 2$s$/2$p$ orbitals of the oxygen $O^{2-}$ anions with the $np$ orbitals of iron cations [43]:

$$V_{zz}^{el} = -\sum_k^6 \tfrac{4}{5}|e|(3\cos\theta_k - 1) \times [\sum_n \langle r_{np}^{-3}\rangle (S_{np}^k)^2] \tag{4A}$$

where the summation over "$k$" extends over all the <Fe-O>$_k$ bonds in the (FeO$_6$) octahedra; $(S_{np}^k)^2 = \{S_{np,s}^2 + S_{np,\sigma}^2 - S_{np,\pi}^2\}$ are the overlap integrals of the $np$ orbitals of the iron with the 2$s$/2$p$ orbitals of the oxygen anions; $\theta_k$ is the angle between the principal $z$ direction of the EFG and line joining the iron cation and the oxygen anion at "$k$" site; $\langle r_{nl}^{-3}\rangle$ refers to the $nl$-wave function of the $Fe^{3+}$ ions closed orbitals ($\langle r_{3p}^{-3}\rangle = 55.93$, and $\langle r_{2p}^{-3}\rangle = 461.8$ in a.u. [43]). The overlap integrals were calculated using 2$s$ and 2$p$ Watson's $O^{2-}$ wave functions in a "3+" stabilizing potential and $np$ Clementi's wave functions for $Fe^{3+}$ cations [43].

At the final stage, to obtain the total EFG $V_{zz}^{tot}$ at the $^{57}$Fe nucleus we corrected for shielding effects produced by the own electrons of the iron ions and external charges:

$$V_{zz}^{tot} = (1-\gamma_\infty)V_{zz}^{lat} + (1-R)V_{zz}^{el}, \qquad (5A)$$

where and $\gamma_\infty = -9.14$ and $R = 0.32$ are Sternheimer factors for these two contributions [30].

## FIGURES CAPTIONS

**Figure 1.** (*Color online*) Schematic crystal (a) and magnetic (b) structure of 3$R$-AgFeO$_2$ (only Fe$^{3+}$ magnetic ions are illustrated). Exchange interactions $J_1$, $J_2$, $J_3$ within the triangular lattice and interplanes exchange interaction $J_0$ are shown.

**Figure 2.** (*Color online*) (a) Temperature dependent magnetic susceptibility ($\chi$) of the 3$R$-AgFeO$_2$ sample. The insert represents an inverse magnetic susceptibility ($\chi^{-1}$) (the solid red lines is the Curie-Weiss law). The positions of arrows correspond to anomalies in the specific heat (at $T_{N2}$ and $T_{N1}$). (b) The temperature dependence of the specific heat ($C_p$). The insert represents an enlarged low temperature region plotting $C_p/T$ with curves taken at $H_{ex}$ = 0; 90 kOe. The arrows indicate the successive magnetic phase transitions at $T_{N2}$ = 9K and $T_{N1}$ = 14K.

**Figure 3.** (*Color online*) (a) $^{57}$Fe Mössbauer spectrum of 3$R$-AgFeO$_2$ recorded at $T$ = 300 K ($T \gg T_{N1}$). The solid red line is the result of simulation of the experimental spectra as described in the text. (b) Theoretical dependences of monopole $V_{zz}^{mon}$, dipole ($V_{xx,yy,zz}^{dip}$) and electronic ($V_{zz}^{el}$) partial contributions to the total EFG (all these contributions include the corresponding Sternheimer factors, *see text*) and resulting quadrupole spitting ($\Delta$) versus the oxygen polarizability ($\alpha_O$). Red circles denote the experimental value of the quadrupole splitting ($\Delta_{exp}$) and the corresponding value of $\alpha_O \approx 0.83$ Å$^3$ (*see text*).

**Figure 4.** (*Color online*) Schematic view of the local crystal structure of 3$R$-AgFeO$_2$ (in hexagonal base) and directions of the principal EFG $\{V_{ii}\}_{i = x,y,z}$ axes, magnetic moments of iron ions ($\mu_{Fe}$), and hyperfine field ($H_{hf}$) at $^{57}$Fe nuclei ($\vartheta$ is the polar angle of the hyperfine field $H_{hf}$ in the principle axes of the EFG tensor.

**Figure 5**. (*Color online*) $^{57}$Fe Mössbauer spectra (experimental hollow dots) of 3$R$-AgFeO$_2$ recorded at the indicated temperatures. Solid red lines are simulation of the experimental spectra as described in the text. The hyperfine field distributions $p(H_{hf})$ resulting from simulation of the spectra are shown on the right.

**Fig. 6.** (*Color online*) Mössbauer determination of the Neel temperature: temperature dependences of the average value of the hyperfine field $\langle H_{hf}\rangle$ and dispersion $D_{p(H)}$ (inset) of the distributions $p(H_{hf})$.

**Fig. 7.** (*Color online*) Reduced hyperfine field $H_{hf}^{(max)}(T)/H_{hf}^{(max)}(4.6)$ as a function of the reduced temperature (red solid line corresponds to fit to the power law given by Eq.(14)). Inset: variation of critical exponents $\beta$ and $\beta^*$ with maximum reduced temperature $t_{max}$ (logarithmic representation).

Blue dashed line corresponds to fit (Eq. (15), see text) and orange dashed line corresponds to the average β value.

**Figure 8**. (*Color online*) (a) $^{57}$Fe Mössbauer spectra (experimental hollow dots) of 3R-AgFeO$_2$ recorded just above the Neel temperature ($T \approx T_{N1}$). Solid lines are simulation of the experimental spectra as the superposition of magnetic (orange area) and paramagnetic (blue line) subspectra (see text). (b) Temperature variation of the fraction of paramagnetic component. Inset: temperature dependence of the relaxation rate ($1/\tau_c$) associated with the two-dimensional spin correlations.

**Figure 9**. (*Color online*) (a) $^{57}$Fe Mössbauer spectra of 3R-AgFeO$_2$ in the $T_{N2} \leq T \leq T_{N1}$ interval fitted with SDW (the red solid line), the sinusoidally modulated SDW (dashed blue line) as described in the text. (b) Resulting shape of the distributions $p_{SDW}(H_{hf})$ (red area); $p_{sin}(H_{hf})$ (dashed blue line) and $p(H_{hf})$ (dark area). The insets (right panels) are the modulations of the hyperfine magnetic field.

**Figure 10.** (*Color online*) Temperature dependences of the angle ($\vartheta$) between the $H_{hf}$ field and the principal component $V_{zz}$ of the EFG. For comparison, it was drown the angle ($\phi$) between the spin direction of SDW and crystal $c_h$ axis taken from Ref. [12]. Inset: temperature dependence of the average value of quadrupole shift ($\varepsilon_Q$).

**Figure 11.** (*Color online*) Temperature dependences of the coefficients of correlation of ($\partial\delta/\partial H_{hf}$) and ($\partial\varepsilon/\partial H_{hf}$) obtained as a result of reconstructing distributions $p(H_{hf})$.

**Figure 12.** *(Color online)* (a) Directions of the principal EFG axes for the iron sites (the angle $\vartheta$ gives the orientation of the hyperfine field $H_{hf}$ in the (*cb*) plane, varying continuously between 0 and $2\pi$; the symbol $H_\parallel$ denotes hyperfine field component on iron along the *c*-axis, while $H_\perp$ stands for the iron hyperfine field component along the *b*-axis) (b) (left site) Polar diagram demonstrating evolution of the spin structure from harmonic distribution (no uniaxial anisotropy, i.e., $D \approx 0$) to squared modulation (very hard uniaxial anisotropy) and (right site) corresponding modulation of the projection $H_{hf}^{(z)}(\vartheta)$ on the *c* axis along the ***q*** direction. (c) Polar diagrams corresponding to the isotropic, $H_\parallel = H_\perp$ (circular polarization) and anisotropic $H_\parallel \neq H_\perp$ (elliptical polarization) hyperfine magnetic fields

**Figure 13**. (*Color online*) $^{57}$Fe Mössbauer spectra at the indicated temperatures fitted using a modulation of the hyperfine interactions as the Fe$^{3+}$ magnetic moment rotates with respect to the principal axis of the EFG tensor, and the anisotropy of the magnetic hyperfine interactions at the Fe$^{3+}$ sites. The hyperfine field distributions $p(H_{hf})$ resulting from simulation of the spectra are shown on the left.

**Figure 14.** (*Color online*) Temperature dependence of the anharmonicity parameter, $H_\parallel$ and $H_\perp$ contributions and the isotropic part ($H_{is}$) of the hyperfine magnetic field at $^{57}$Fe nuclei extracted from least-squares fits of the Mössbauer spectra.

**Figure 15.** (*Color online*) Plot of the components $H_{hf}^{(z)}(\vartheta)$ and $H_{hf}^{(x)}(\vartheta)$ according to the distorted cycloidal model ($<m> = 0.78$) proposed for 3R-AgFeO$_2$ in the range $T < T_{N2}$. These functions are compared with sinusoidal dependence $H_{\text{hf}}^{(z)}(\vartheta) \sim \sin\vartheta$ for undistorted circular cycloid (dashed red line).

**Figure 16**. (*Color online*) Schematic representation of different contributions ($\Delta h_{STHF}$, $\Delta h_{\text{dir}}$, $\Delta H_{\text{red}}$) to the $H_{\text{hf}}$ value for the ferrites: (564 kOe ←) LaFeO$_3$, (515 kOe ←) CuFeO$_2$ ($T << T_{\text{N2}}$), and (484 kOe ←) 3R-AgFeO$_2$ ($T << T_{\text{N2}}$). The dashed red line represents the calculated sum of contributions $H_F + H_{\text{cov}} \approx 494$ kOe (*see text*), which was considered the same for the three ferrites. The figure also shows the local magnetic structures of the ferrites (green shaded areas correspond to spins giving the opposite in sign contributions ($\Delta h_{STHF}$ or $\Delta h_{\text{dir}}$) to the total $H_{\text{hf}}$ value, *see text*).

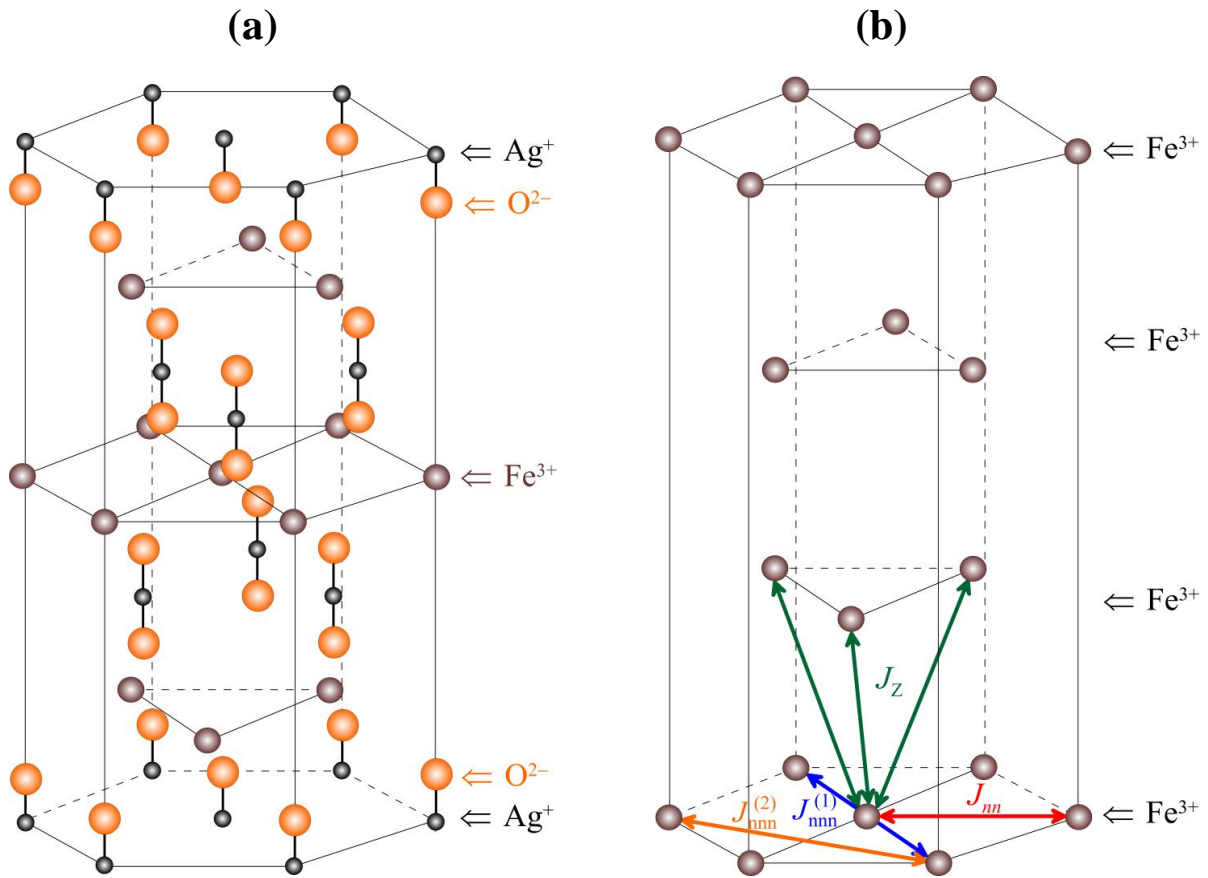

**Figure 1.** (*Color online*) Schematic crystal (a) and magnetic (b) structure of 3$R$-AgFeO$_2$ (only Fe$^{3+}$ magnetic ions are illustrated). Exchange interactions $J_1$, $J_2$, $J_3$ within the triangular lattice and interplanes exchange interaction $J_0$ are shown.

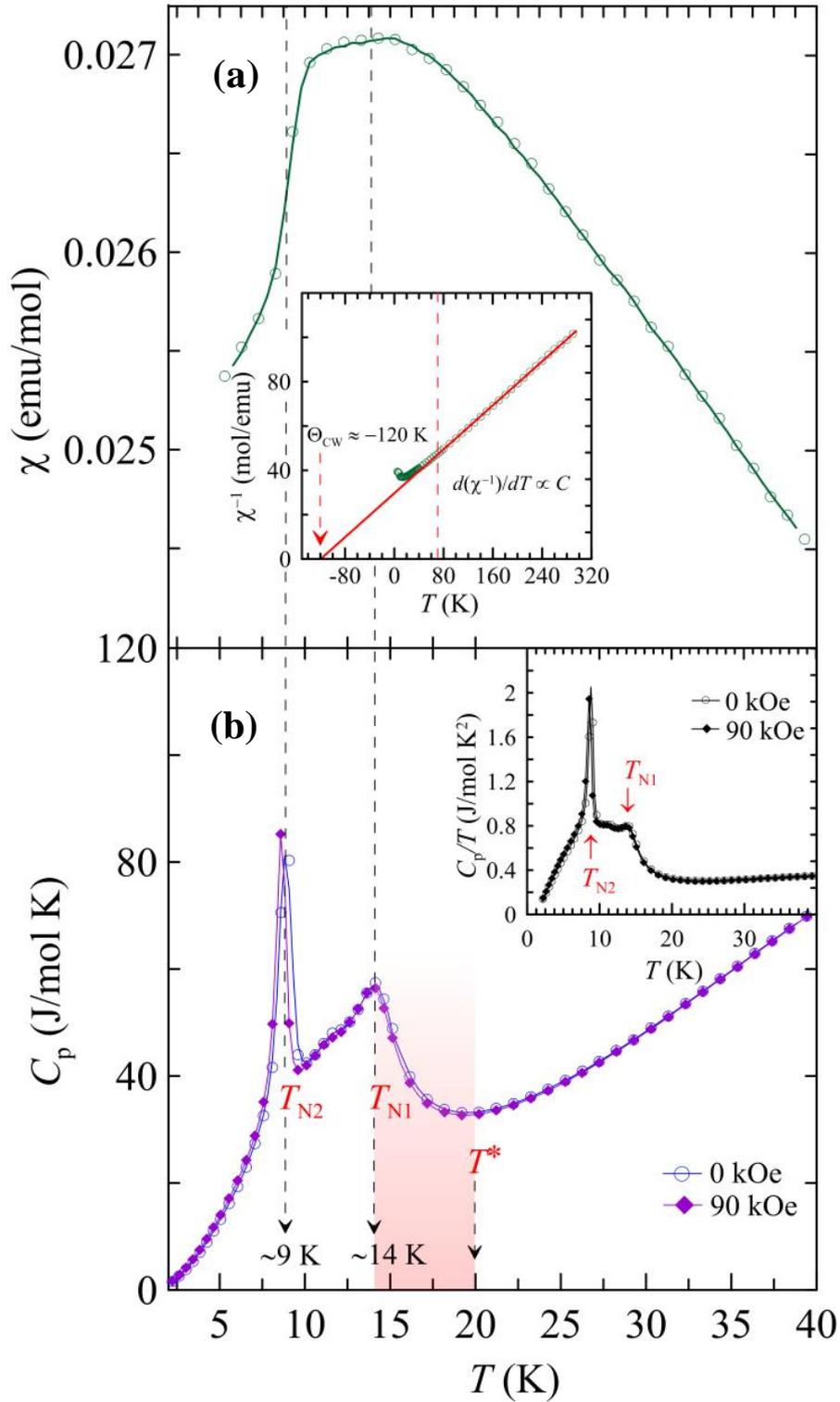

**Figure 2.** (*Color online*) (a) Temperature dependent magnetic susceptibility ($\chi$) of the 3$R$-AgFeO$_2$ sample. The insert represents an inverse magnetic susceptibility ($\chi^{-1}$) (the solid red lines is the Curie-Weiss law). The positions of arrows correspond to anomalies in the specific heat (at $T_{N2}$ and $T_{N1}$). (b) The temperature dependence of the specific heat ($C_p$). The insert represents an enlarged low temperature region plotting $C_p/T$ with curves taken at $H_{ex} = 0$; 90 kOe. The arrows indicate the successive magnetic phase transitions at $T_{N2} = 9$K and $T_{N1} = 14$K.

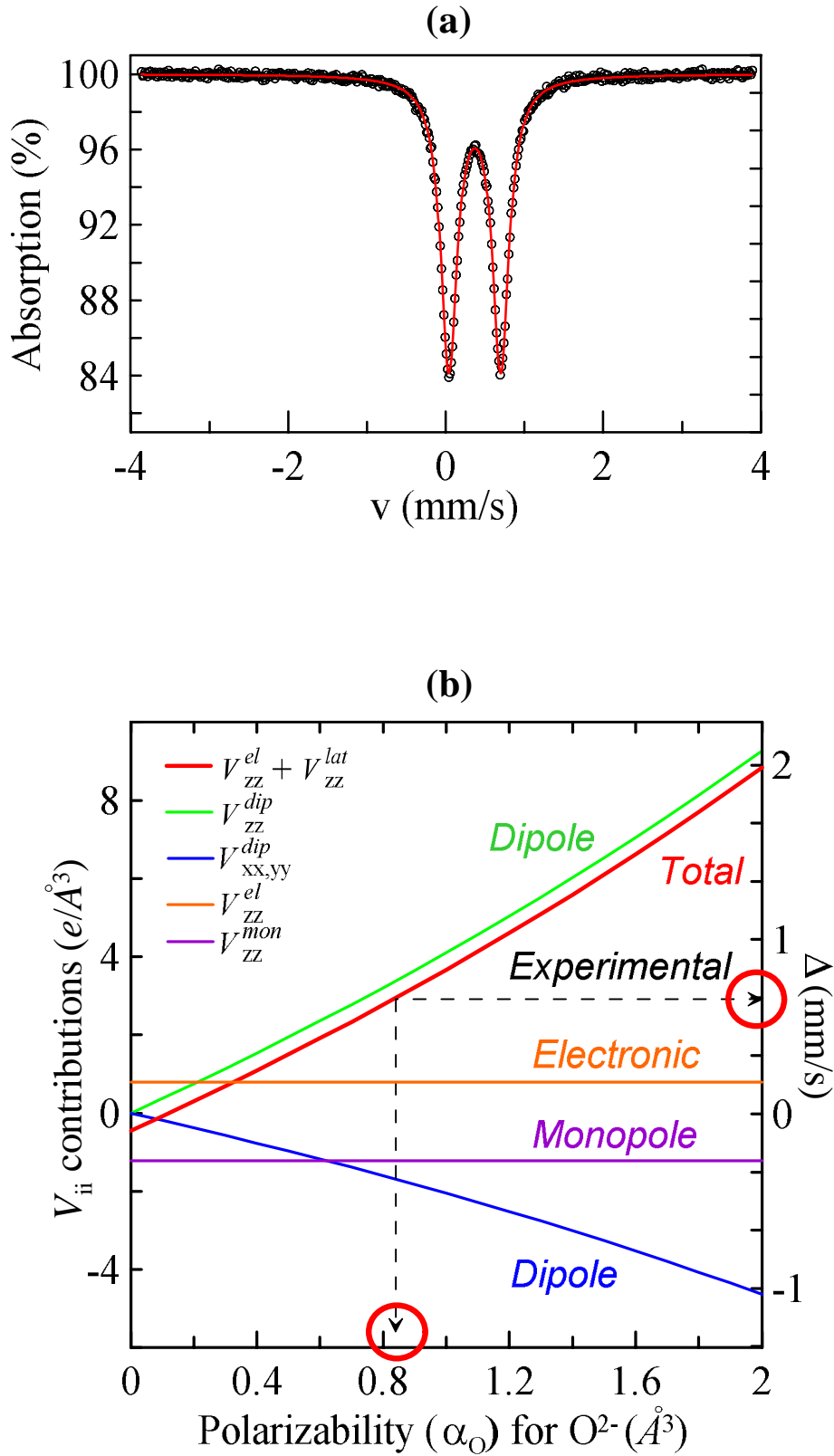

**Figure 3.** (*Color online*) (a) $^{57}$Fe Mössbauer spectrum of 3$R$-AgFeO$_2$ recorded at $T = 300$ K ($T \gg T_{N1}$). The solid red line is the result of simulation of the experimental spectra as described in the text. (b) Theoretical dependences of monopole $V_{zz}^{mon}$, dipole ($V_{xx,yy,zz}^{dip}$) and electronic ($V_{zz}^{el}$) partial contributions to the total EFG (all these contributions include the corresponding Sternheimer factors, *see text*) and resulting quadrupole spitting ($\Delta$) versus the oxygen polarizability ($\alpha_O$). Red circles denote the experimental value of the quadrupole splitting ($\Delta_{exp}$) and the corresponding value of $\alpha_O \approx 0.83$ Å$^3$ (*see text*).

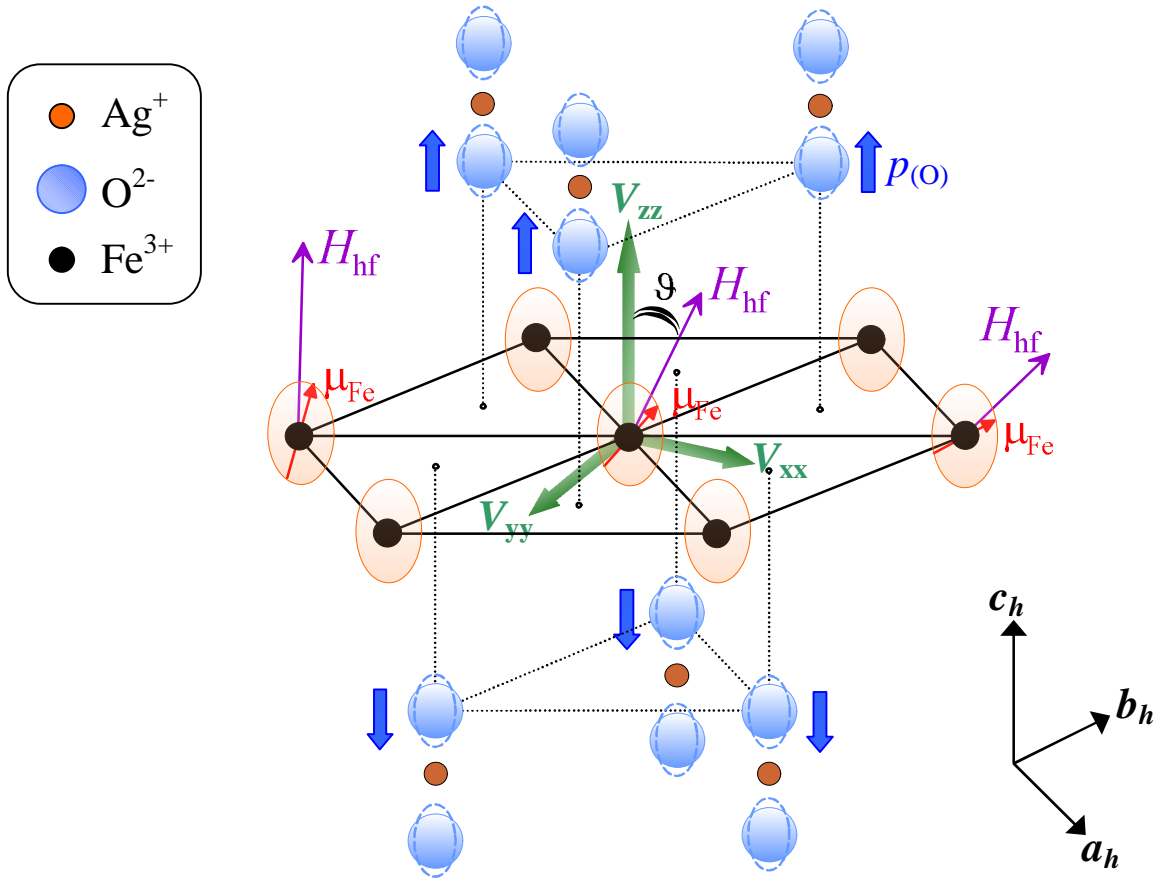

**Figure 4.** (*Color online*) Schematic view of the local crystal structure of 3$R$-AgFeO$_2$ (in hexagonal base) and directions of the principal EFG $\{V_{ii}\}_{i=x,y,z}$ axes, magnetic moments of iron ions ($\mu_{Fe}$), and hyperfine field ($H_{hf}$) at $^{57}$Fe nuclei ($\vartheta$ is the polar angle of the hyperfine field $H_{hf}$ in the principle axes of the EFG tensor.

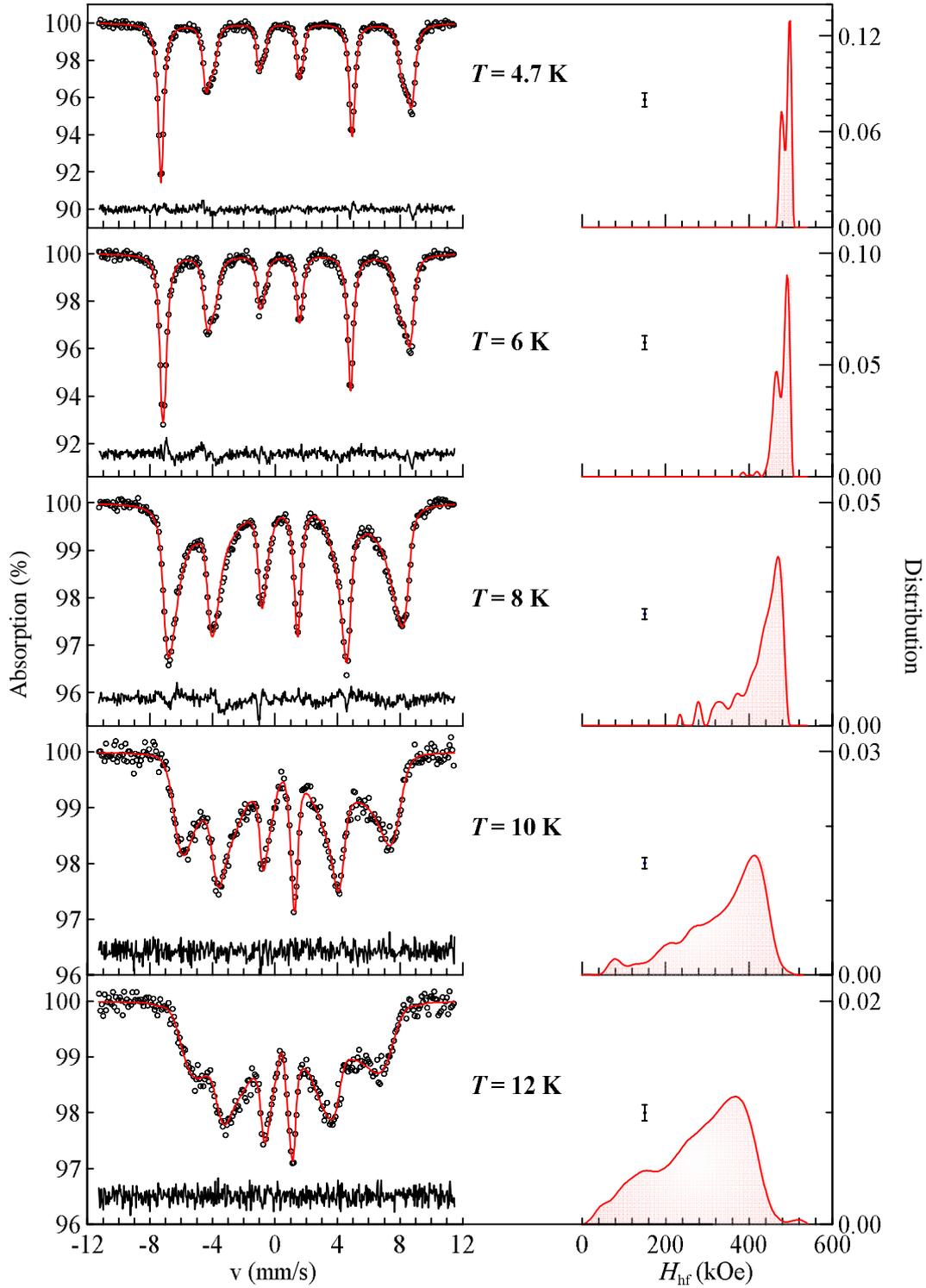

**Figure 5**. (*Color online*) $^{57}$Fe Mössbauer spectra (experimental hollow dots) of 3*R*-AgFeO$_2$ recorded at the indicated temperatures. Solid red lines are simulation of the experimental spectra as described in the text. The hyperfine field distributions $p(H_{hf})$ resulting from simulation of the spectra are shown on the right.

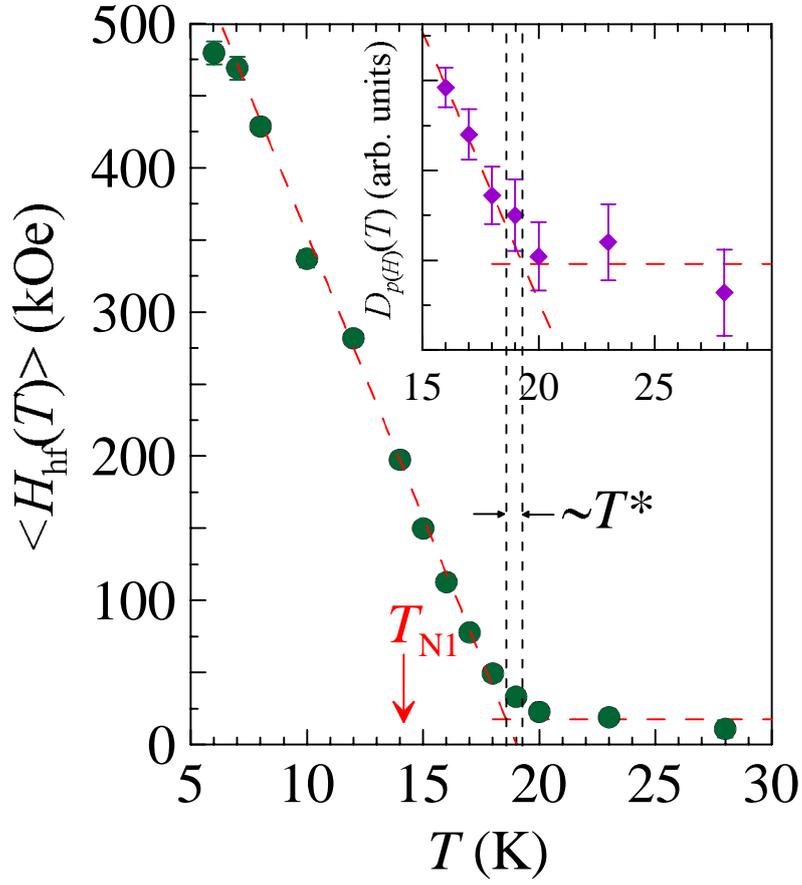

**Fig. 6.** (*Color online*) Mössbauer determination of the Neel temperature: temperature dependences of the average value of the hyperfine field $\langle H_{hf}\rangle$ and dispersion $D_{p(H)}$ (inset) of the distributions $p(H_{hf})$.

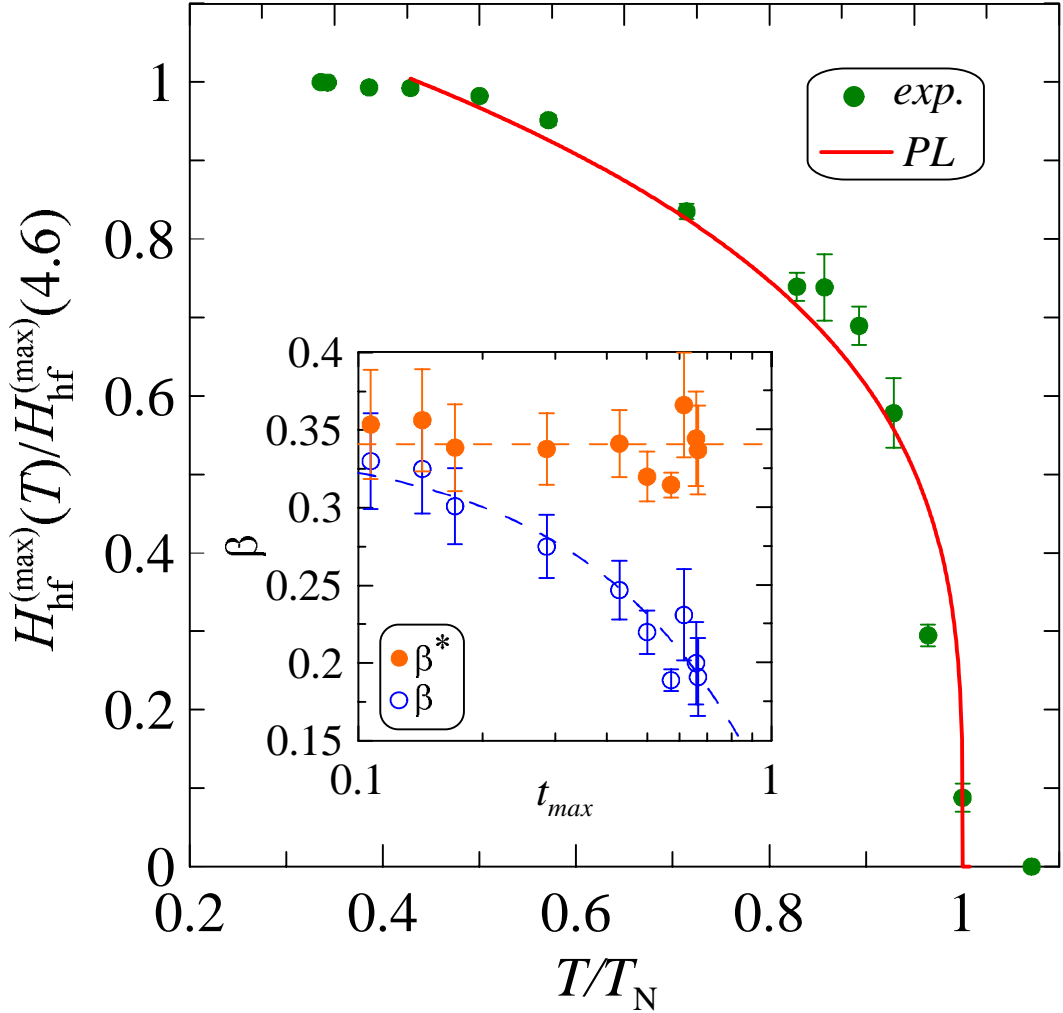

**Fig. 7.** (*Color online*) Reduced hyperfine field $H_{\text{hf}}^{(\text{max})}(T)/H_{\text{hf}}^{(\text{max})}(4.6)$ as a function of the reduced temperature (red solid line corresponds to fit to the power law given by Eq.(14)). Inset: variation of critical exponents $\beta$ and $\beta^*$ with maximum reduced temperature $t_{max}$ (logarithmic representation). Blue dashed line corresponds to fit (Eq. (15), see text) and orange dashed line corresponds to the average $\beta$ value.

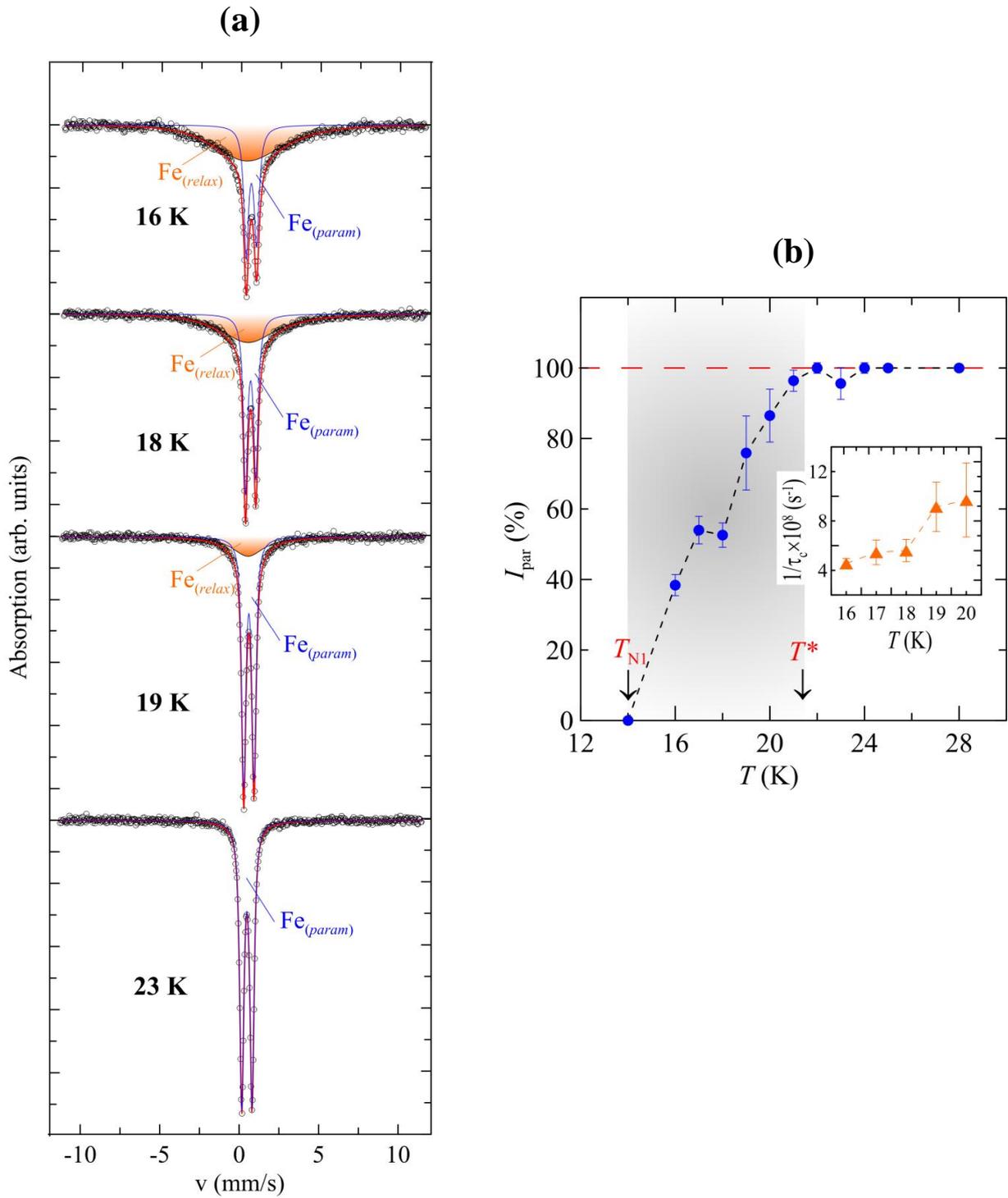

**Figure 8**. (*Color online*) (a) $^{57}$Fe Mössbauer spectra (experimental hollow dots) of 3$R$-AgFeO$_2$ recorded just above the Neel temperature ($T \approx T_{N1}$). Solid lines are simulation of the experimental spectra as the superposition of magnetic (orange area) and paramagnetic (blue line) subspectra (see text). (b) Temperature variation of the fraction of paramagnetic component. Inset: temperature dependence of the relaxation rate ($1/\tau_c$) associated with the two-dimensional spin correlations.

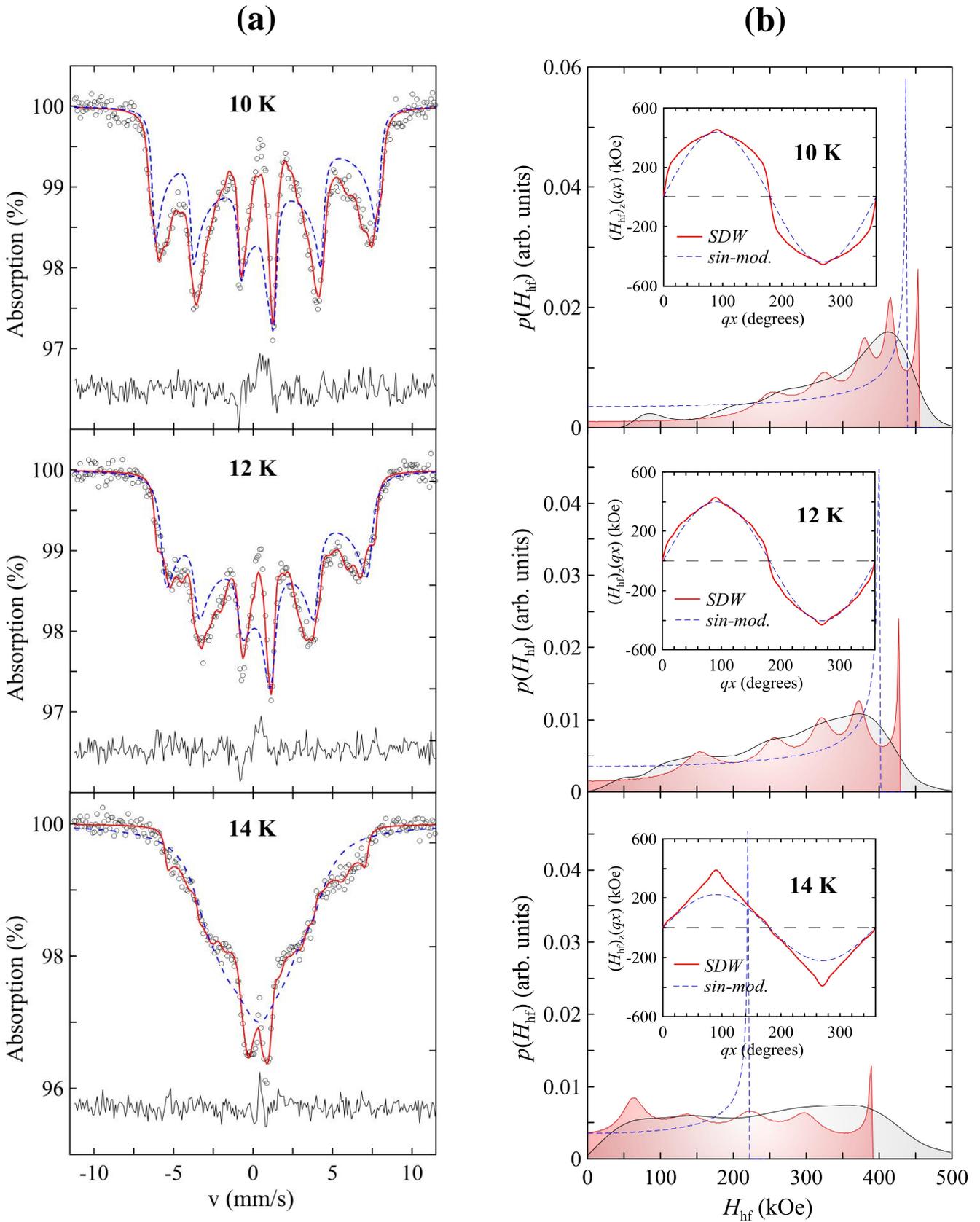

**Figure 9**. (*Color online*) (a) $^{57}$Fe Mössbauer spectra of 3R-AgFeO$_2$ in the $T_{N2} \leq T \leq T_{N1}$ interval fitted with SDW (the red solid line), the sinusoidally modulated SDW (dashed blue line) as described in the text. (b) Resulting shape of the distributions $p_{SDW}(H_{hf})$ (red area); $p_{sin}(H_{hf})$ (dashed blue line) and $p(H_{hf})$ (dark area). The insets (right panels) are the modulations of the hyperfine magnetic field.

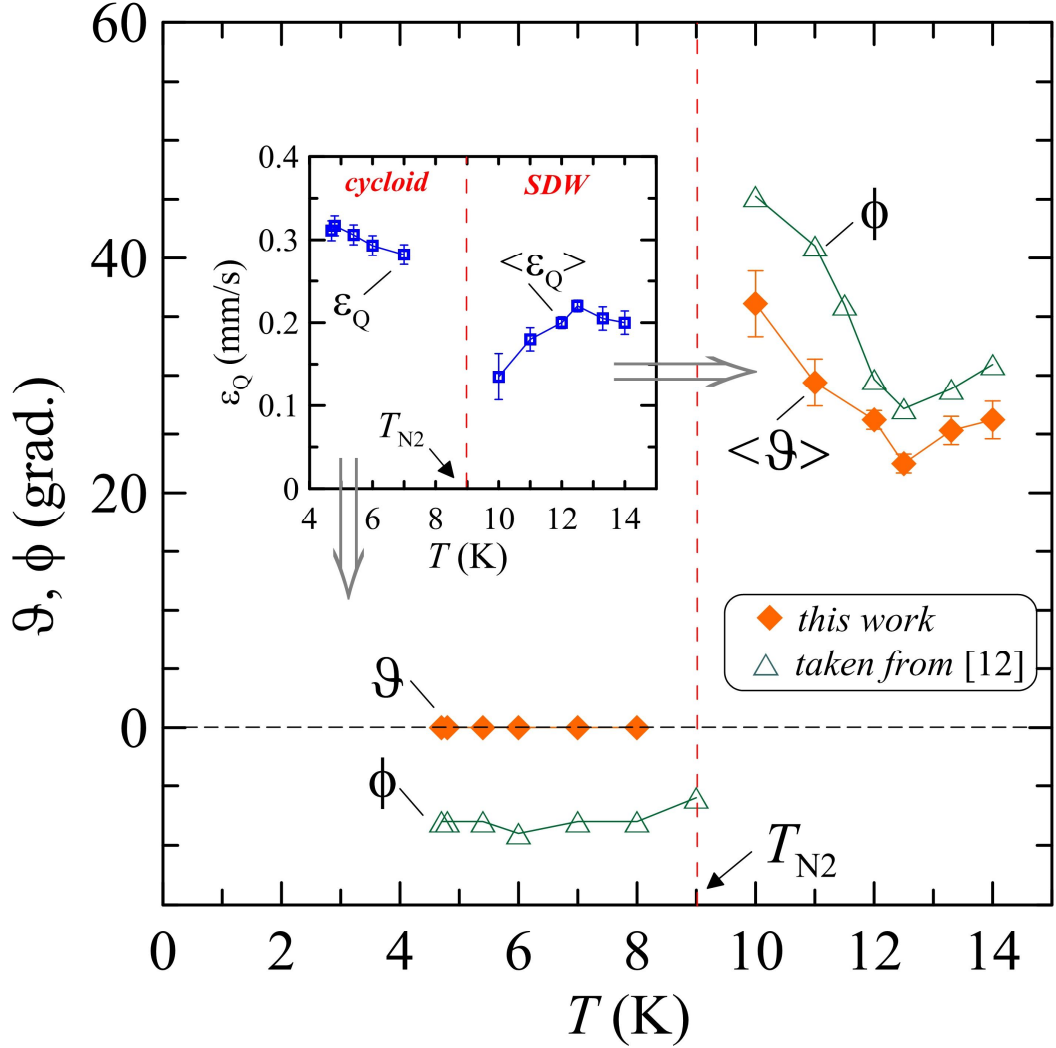

**Figure 10.** (*Color online*) Temperature dependences of the angle ($\vartheta$) between the $H_{hf}$ field and the principal component $V_{zz}$ of the EFG. For comparison, it was drown the angle ($\phi$) between the spin direction of SDW and crystal $c_h$ axis taken from Ref. [12]. Inset: temperature dependence of the average value of quadrupole shift ($\varepsilon_Q$).

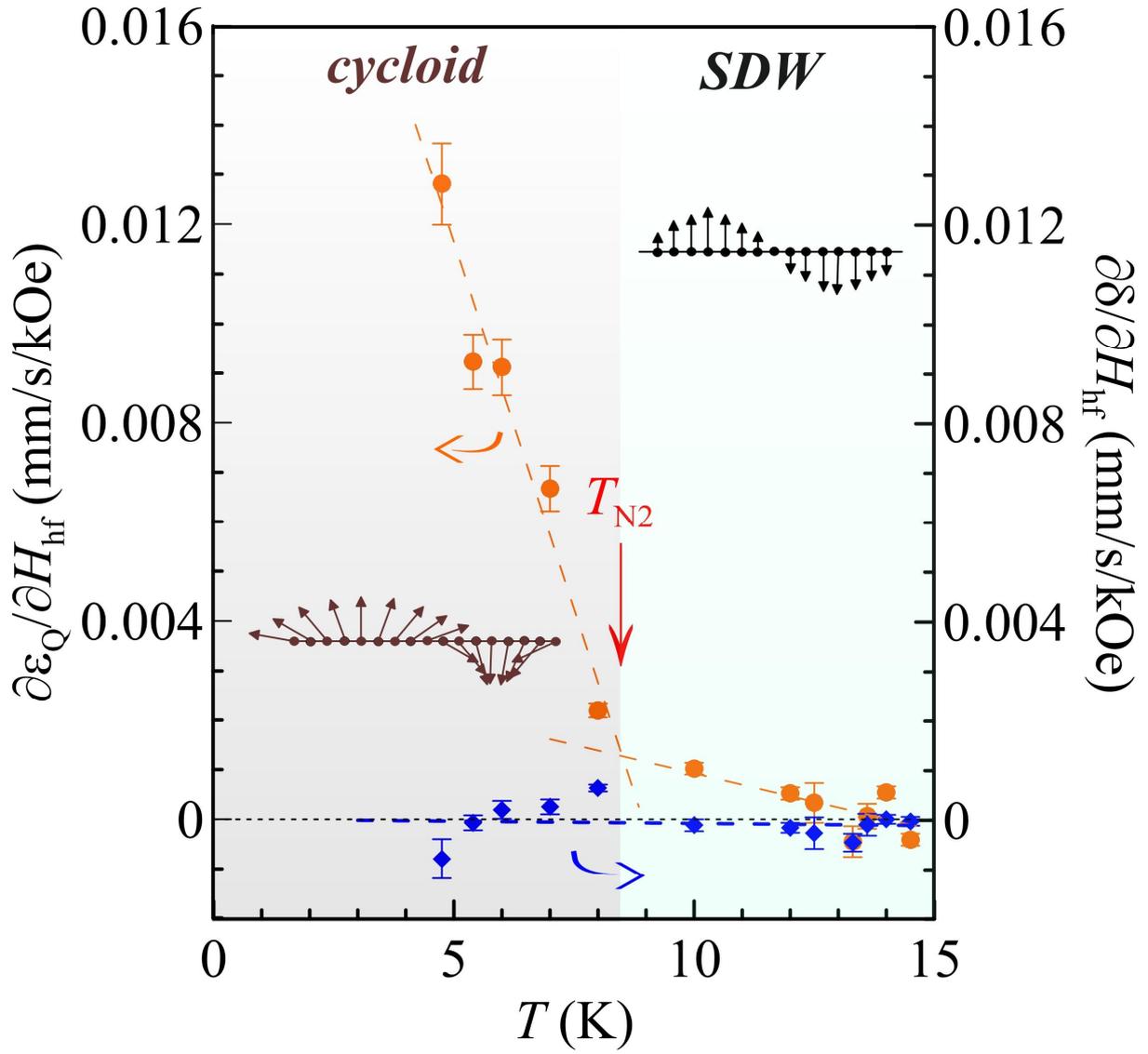

**Figure 11**. (*Color online*) Temperature dependences of the coefficients of correlation of ($\partial\delta/\partial H_{hf}$) and ($\partial\varepsilon/\partial H_{hf}$) obtained as a result of reconstructing distributions $p(H_{hf})$.

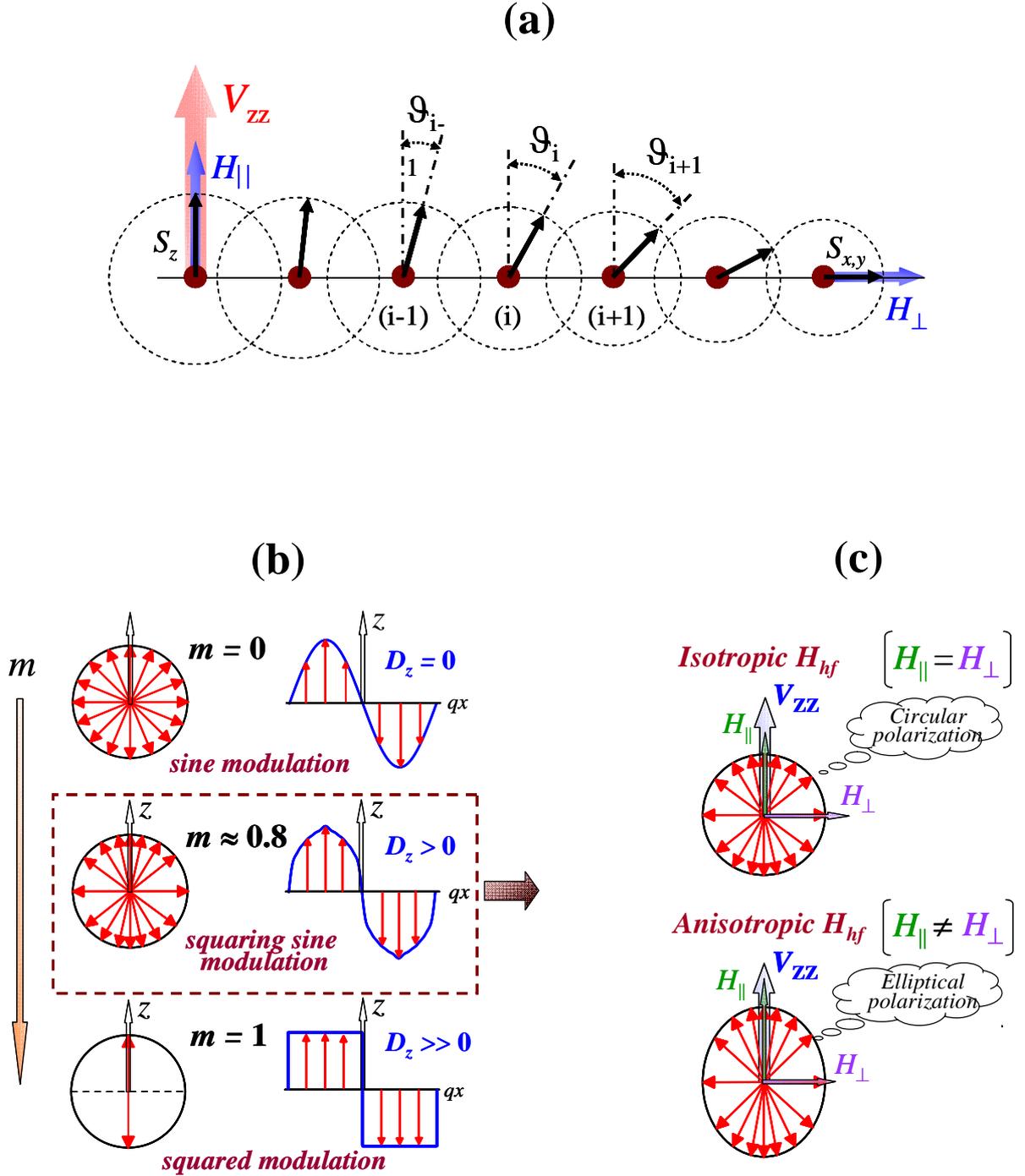

**Figure 12.** *(Color online)* (a) Directions of the principal EFG axes for the iron sites (the angle $\vartheta$ gives the orientation of the hyperfine field $H_{hf}$ in the ($cb$) plane, varying continuously between 0 and $2\pi$; the symbol $H_{\parallel}$ denotes hyperfine field component on iron along the $c$-axis, while $H_{\perp}$ stands for the iron hyperfine field component along the $b$-axis) (b) (left site) Polar diagram demonstrating evolution of the spin structure from harmonic distribution (no uniaxial anisotropy, i.e., $D \approx 0$) to squared modulation (very hard uniaxial anisotropy) and (right site) corresponding modulation of the projection $H_{hf}^{(z)}(\vartheta)$ on the $c$ axis along the $q$ direction. (c) Polar diagrams corresponding to the isotropic, $H_{\parallel} = H_{\perp}$ (circular polarization) and anisotropic $H_{\parallel} \neq H_{\perp}$ (elliptical polarization) hyperfine magnetic fields

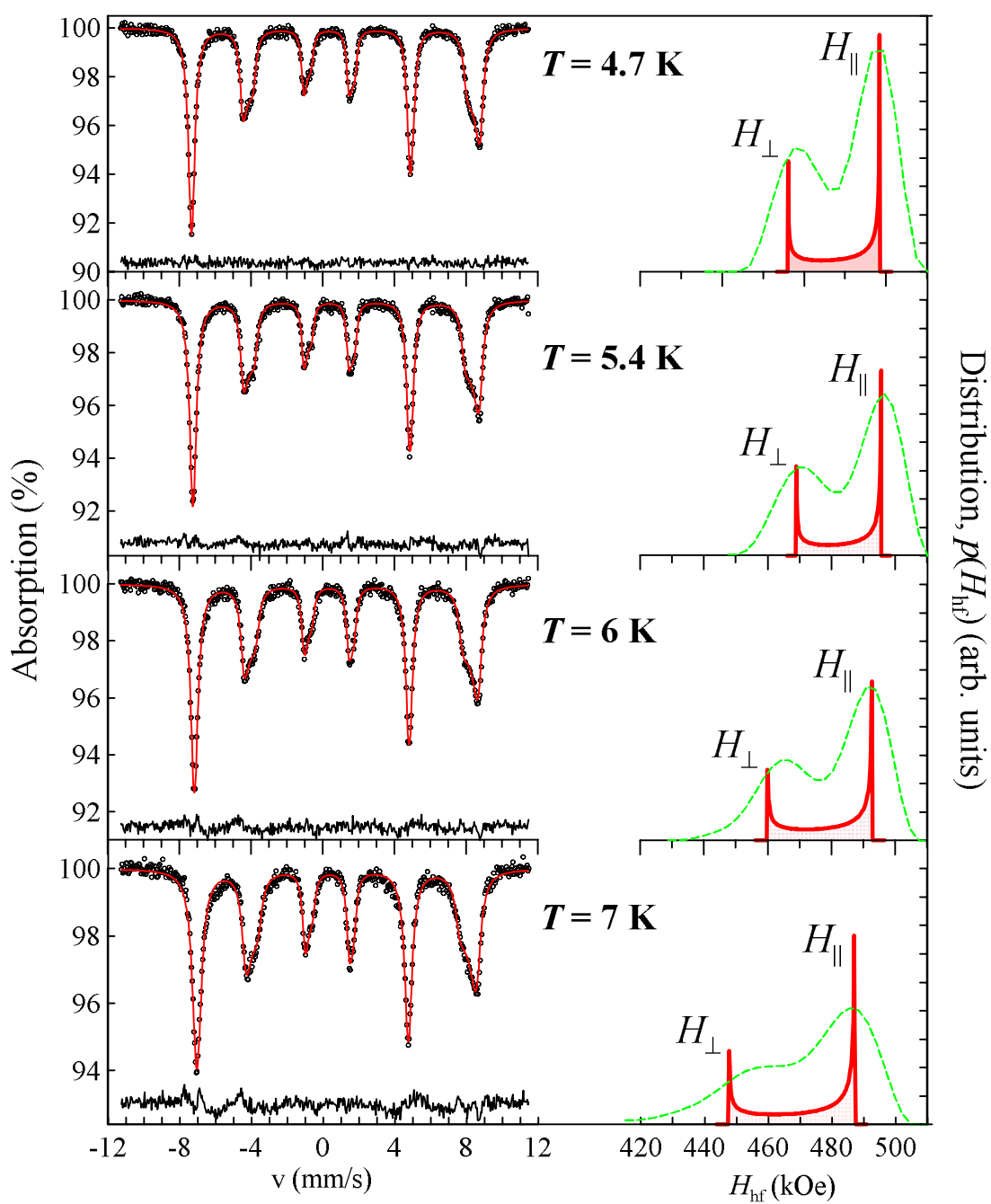

**Figure 13**. (*Color online*) $^{57}$Fe Mössbauer spectra at the indicated temperatures fitted using a modulation of the hyperfine interactions as the $Fe^{3+}$ magnetic moment rotates with respect to the principal axis of the EFG tensor, and the anisotropy of the magnetic hyperfine interactions at the $Fe^{3+}$ sites. The hyperfine field distributions $p(H_{hf})$ resulting from simulation of the spectra are shown on the left.

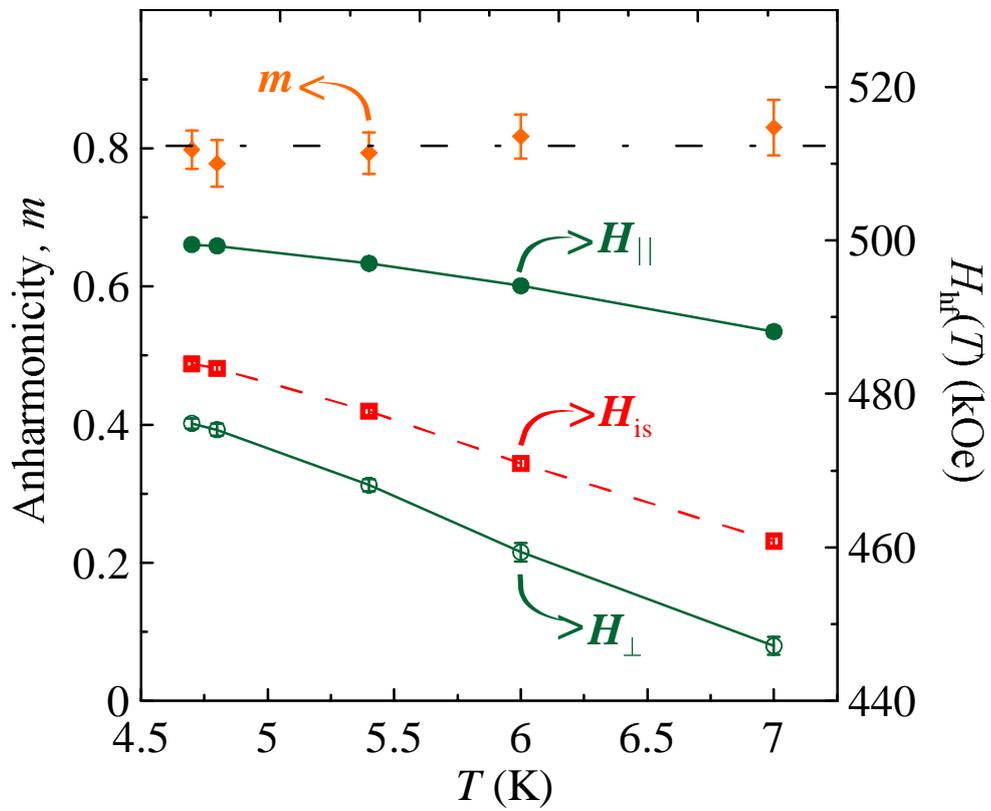

**Figure 14.** (*Color online*) Temperature dependence of the anharmonicity parameter, $H_\parallel$ and $H_\perp$ contributions and the isotropic part ($H_{is}$) of the hyperfine magnetic field at $^{57}$Fe nuclei extracted from least-squares fits of the Mössbauer spectra.

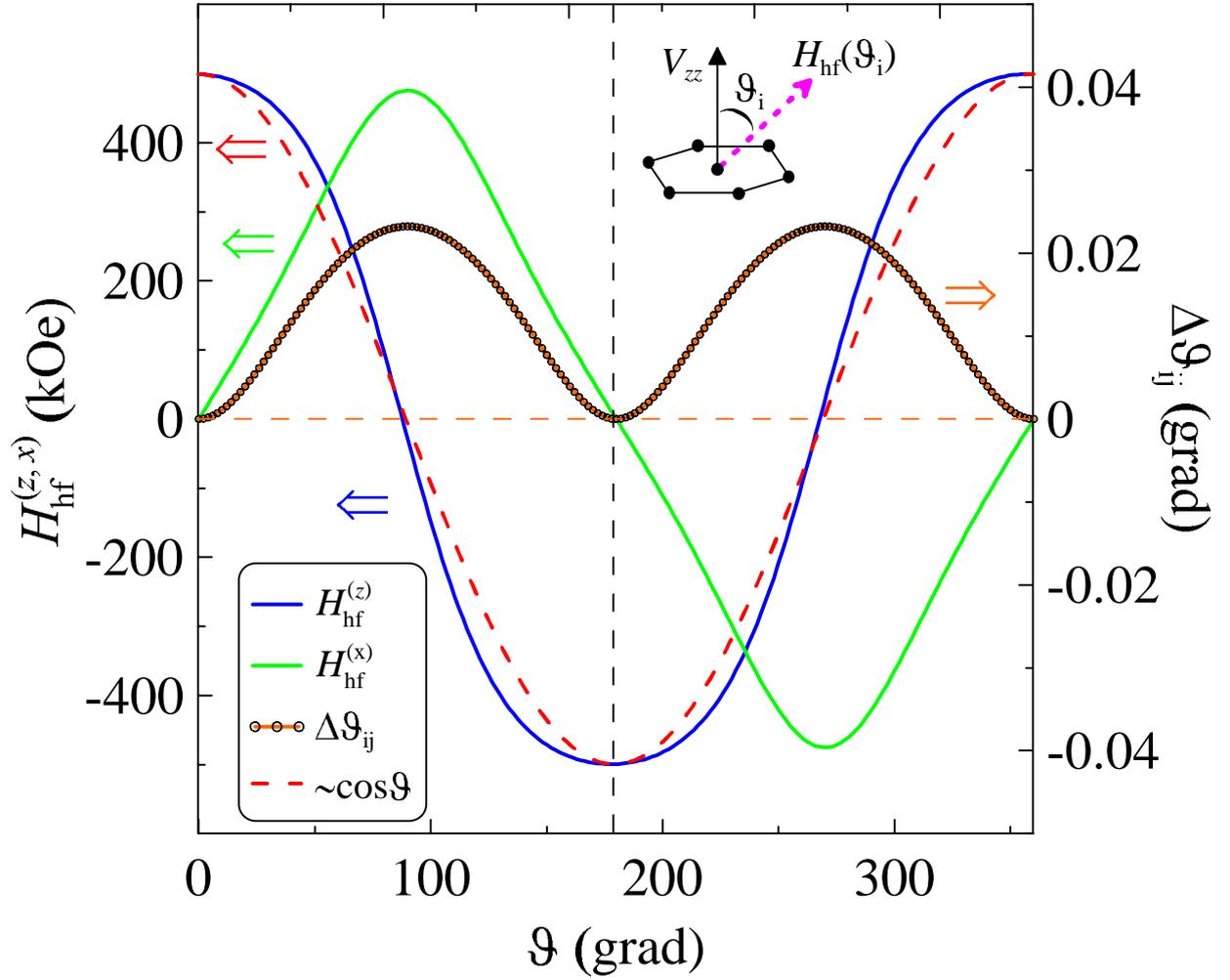

**Figure 15.** (*Color online*) Plot of the components $H_{hf}^{(z)}(\vartheta)$ and $H_{hf}^{(x)}(\vartheta)$ according to the distorted cycloidal model (<$m$> = 0.78) proposed for 3$R$-AgFeO$_2$ in the range $T < T_{N2}$. These functions are compared with sinusoidal dependence $H_{hf}^{(z)}(\vartheta) \sim \sin\vartheta$ for undistorted circular cycloid (dashed red line).

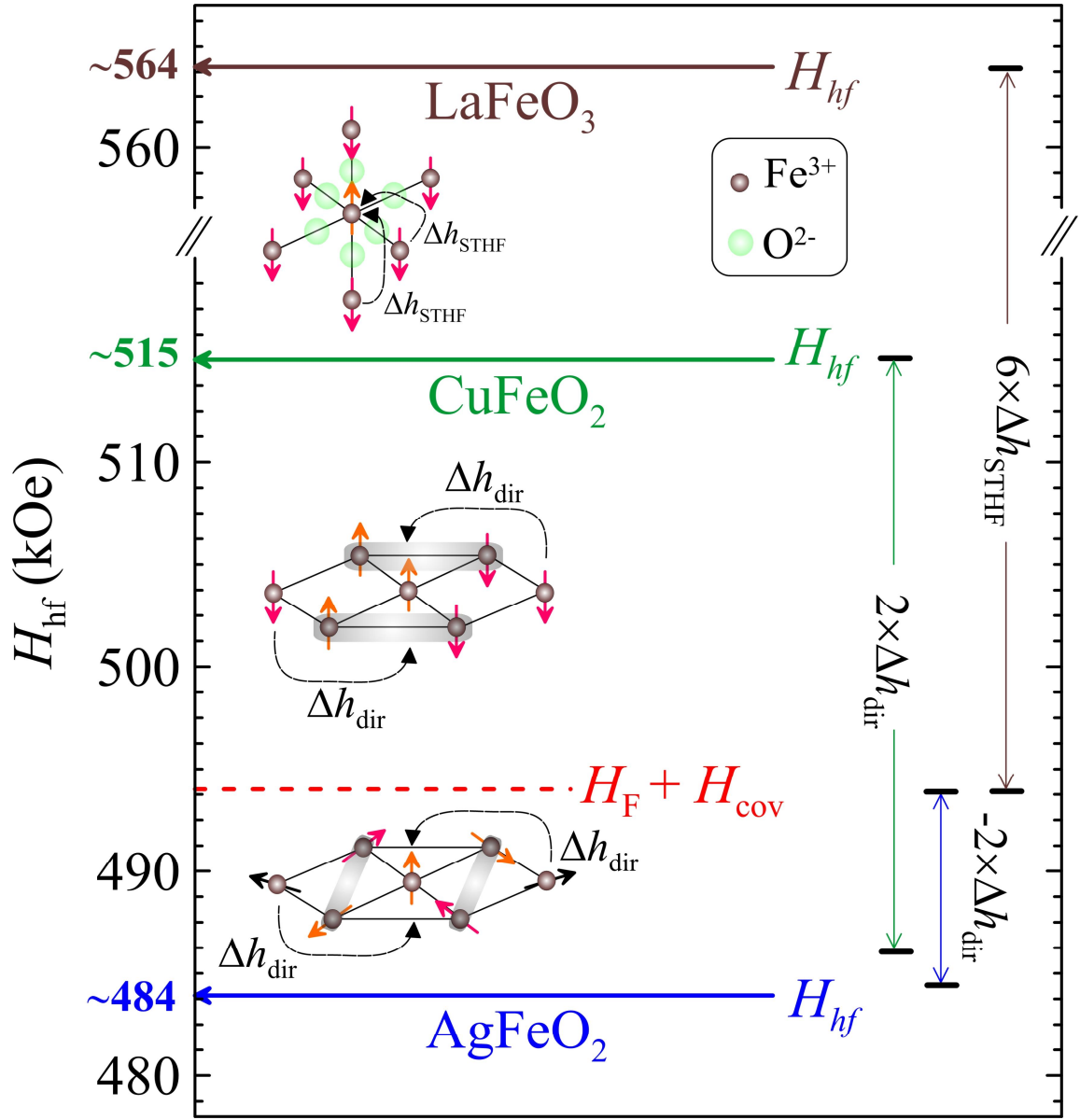

**Figure 16.** (*Color online*) Schematic representation of different contributions ($\Delta h_{STHF}$, $\Delta h_{dir}$, $\Delta H_{red}$) to the $H_{hf}$ value for the ferrites: (564 kOe ←) LaFeO$_3$, (515 kOe ←) CuFeO$_2$ ($T \ll T_{N2}$), and (484 kOe ←) 3R-AgFeO$_2$ ($T \ll T_{N2}$). The dashed red line represents the calculated sum of contributions $H_F + H_{cov} \approx 494$ kOe (*see text*), which was considered the same for the three ferrites. The figure also shows the local magnetic structures of the ferrites (green shaded areas correspond to spins giving the opposite in sign contributions ($\Delta h_{STHF}$ or $\Delta h_{dir}$) to the total $H_{hf}$ value, *see text*).